\documentclass[12pt, fleqn]{article}
\usepackage[cp1251]{inputenc}
\usepackage{latexsym,amsfonts,amssymb}
\usepackage{graphicx}

\usepackage{amsbsy}
\usepackage{amsmath}
\usepackage{epsf}

\newtheorem{theo} {Theorem}
\newcommand{\bteo}{\begin{theo}}  
\newcommand{\et}{\end{theo}}
\newtheorem{Remark}{Remark}
\newtheorem{definition}{Definition}
\newcommand{\bd}{\begin{displaymath}}
\newcommand{\ed}{\end{displaymath}}

\newcommand{\lf}{\left}
\newcommand{\rg}{\right}
\newcommand{\be} {\begin{equation}}
\newcommand{\ee} {\end{equation}}
\newcommand{\ba} {\begin{array}}
\newcommand{\ea} {\end{array}}
\newcommand{\bea}{\begin{eqnarray}}
\newcommand{\eea} {\end{eqnarray}}
\newcommand{\p} {\partial}

\begin{document}

\begin{center}
 {\Large \bf Symmetries and  exact solutions of the diffusive Holling--Tanner prey-predator model}

\medskip

{\bf Roman Cherniha $^{a,b,}$\footnote{\small  Corresponding author.
E-mail: r.m.cherniha@gmail.com; roman.cherniha1@nottingham.ac.uk}
and Vasyl' Davydovych$^{a}$}

$^{a}$ \quad Institute of Mathematics,  National Academy of Sciences
of Ukraine, \\
 3, Tereshchenkivs'ka Street, Kyiv 01004, Ukraine.\\
 $^{b}$ \quad School of Mathematical Sciences, University of Nottingham,\\
  University Park, Nottingham NG7 2RD, UK
\end{center}

\begin{abstract} We consider  the classical Holling--Tanner model extended
on 1D space by introducing the diffusion term.
 Making a reasonable simplification, the diffusive Holling--Tanner system
 is studied by means of symmetry based methods. Lie and
   $Q$-conditional (nonclassical) symmetries are identified.
   The symmetries obtained are applied for finding  a wide range of exact solutions, their properties are studied  and
   a  possible biological interpretation is proposed. 3D plots of the most interesting solutions are drown as well.
\end{abstract}

\section{Introduction} \label{sec-1}

It is well-known that different types of the interaction between
species (cells, chemicals, etc.) occur in nature. There are many
mathematical models describing those in time and/or space (see,
e.g., the well-known books
\cite{aris-75I,britton,fife-79,ku-na-ei-16,mur2, mur2002,
mur2003,okubo}).
 One of the most common interaction between species is one between preys and
 predators. The first mathematical model for such a type of interaction was independently developed
 by  Lotka \cite{lot} and  Volterra \cite{vol}.
 The  model is based on a system of
 ordinary differential equations
  (ODEs) involving  quadratic nonlinearities and is called the
  Lotka--Volterra prey-predator model. Although this model correctly  predicts
  (at least qualitatively) the time evolution of prey and predator
  numbers,  there are some other models used for the same purposes.
  One of the most important  is the Holling--Tanner prey-predator model \cite{hanski91,hanski93,tanner,wollkind}, which
  is much more sophisticated and seems to be more adequate  than the Lotka--Volterra model
   (see, e.g., a note on P.79 in \cite{mur2}).

   The  Holling--Tanner model reads as
   \begin{equation}\label{0}\begin{array}{l}\frac{dU}{dt}=rU\Big(1- \frac{U}{K}\Big)- \frac{q\,UV}{U+A},\medskip\\
\frac{dV}{dt}=sV\Big(1- \frac{V}{hU}\Big),\end{array}\end{equation}
where unknown functions $U(t)$ and $V(t)$ represent
 the numbers of
preys and predators at time $t$, respectively;  $r, \ K, \ q, \ A, \
s$ and $h$ are positive parameters and  their detailed
 description can be found  in \cite{hanski91} and \cite{tanner}.
 Assuming that a given  ecological system produces unbounded
 amount of food for preys, i.e. $K \to \infty$, we may skip the
 quadratic term in the first equation of  (\ref{0}).
 On the other hand, the  diffusion  of preys and predators in space
 can be taking into account \cite{arancibia-21,alaoui-18,chen-12,Qi-16}. As a result, the following system of
 reaction-diffusion equations in 1D approximation is obtained
\be\label{0-1}\ba{l}
u_t = d_1u_{xx}+ru - \frac{q\,uv}{u+A}, \medskip\\
v_t = d_2v_{xx}+ sv\Big(1- \frac{v}{hu}\Big), \ea \ee where unknown
functions $u(t,x)$ and $v(t,x)$ represent the concentrations of
preys and predators, respectively  (the lower subscripts  $t$ and
$x$ denote differentiation with respect to these variables).
Assuming that both diffusivities $d_1$ and $d_2$ are positive
constants (generally speaking, the case  $d_1d_2=0$ can also occur,
but we prefer to discuss this special case elsewhere) and $r$ and
$h$ are also positive (otherwise the model loses its biological
meaning), one can essentially reduce the number of parameters in the
nonlinear system  (\ref{0-1}). In fact,
 using the following substitution
\[t \rightarrow \frac{t}{r}, \ x \rightarrow \sqrt{\frac{d_1}{r}}\,x, \ u\rightarrow u, \ v \rightarrow hv\]
and introducing new notation \[d=\frac{d_2}{d_1}, \ R=\frac{hq}{r},
\ S=\frac{s}{r},\]
 we arrive at the system
\be\label{1-1}\ba{l}
u_t = u_{xx}+u(1-\frac{R\,v}{u+A}), \medskip\\
v_t = dv_{xx}+ Sv\lf(1-\frac{v}{u}\rg), \ea\ee
 where $A$ is arbitrary nonnegative constant,
  $R$ and $S$ are arbitrary positive
constants,  $d>0$ is a nondimensional
 diffusion coefficient.
 In what follows the nonlinear system of partial differential
equations (PDEs) (\ref{1-1}) is called the
 diffusive  Holling--Tanner model (the DHT model) and is the  main object of
 investigation in this work. The DHT model is
  studied by means of two most effective symmetry based methods -- classical Lie
    and $Q$-conditional (nonclassical) symmetry methods. Both methods are
     well-known and described in many excellent works, including the most recent books
      \cite{arrigo-15,bl-anco-10,ch-se-pl-book}. Notably the monograph \cite{ch-dav-book}
  is  devoted to conditional symmetries of reaction-diffusion systems and
  the DHT system (\ref{1-1}) belongs to this class of  systems.
  There are several recent studies devoted to investigation of  systems of
   reaction-diffusion equations by symmetry based methods, in particular \cite{bro-ch-goa-22,ch-dav-AAM-22,ji-16,torrisi-21,torrisi-23}.

  Although the DHT model was studies extensively by different mathematical techniques
   (see \cite{arancibia-21,alaoui-18,chen-12,Qi-16} and the references cited therein), to the best of our knowledge,
    Lie and conditional  symmetries and exact solutions of
     this model were still  unknown. Thus, the main aim of the present  study is to eliminate this gap.

In Section~\ref{sec-2}, all possible  Lie symmetries of the DHT
prey-predator model are identified. It turns out that  system
(\ref{1-1}), depending on values of  parameters, admits essentially
different Lie algebras   of invariance. In Section~\ref{sec-3}, a
wide range of exact solutions of the nonlinear  system in question
are constructed by means of Lie symmetries derived in
Section~\ref{sec-2}. In Section~\ref{sec-4}, highly nontrivial
$Q$-conditional (nonclassical) symmetries  are found using the
notion of the $Q$-conditional symmetry of the first type
\cite{ch-2010}. In Section~\ref{sec-5}, further classes of  exact
solutions are derived by applying  the $Q$-conditional symmetries
obtained. A possible  biological interpretation of some solutions is
discussed in Sections~\ref{sec-3} and~\ref{sec-5} as well. Finally,
we discuss the results obtained and  present some conclusions  in
the last section.


\section {Lie  symmetry classification and reduction to ODEs} \label{sec-2}

Obviously,  the DHT system (\ref{1-1}) with arbitrary coefficients
is invariant under the  two-dimensional Lie algebra generated by the
following operators\,: \be\label{2-1}P_t=\p_t, \quad P_x=\p_x.\ee
This algebra is usually called the principal (trivial) algebra
\cite{ch-se-pl-book}.

It turns out that there are several cases when the DHT system
(\ref{1-1})  admits  nontrivial extensions of the Lie symmetry
(\ref{2-1}), depending on the parameters $A, \ R, \ S$ and $d$.
All possible extensions  have been  identified and are presented as
follows.

\begin{theo}\label{th-1} The DHT system (\ref{1-1}) with arbitrary parameters $A, \ R, \ S$ and $d$ is invariant with respect
to two-dimensional Lie algebra of invariance (\ref{2-1}) . This
system admits three-dimensional or  higher-dimensional
 Lie algebras of invariance if and only if its parameters  have the forms listed in
Table~\ref{tab1}. \end{theo}
\begin{table}[h!]
\caption{Lie symmetries of the  DHT system (\ref{1-1})}\medskip
\label{tab1}       
\begin{tabular}{|c|c|c|}
\hline  & Restrictions  &  Lie algebra of invariance  \\
 \hline && \\
 1 & $A=0$  & $P_t, \  P_x, \  I=u\p_u+v\p_v$ \\ \hline &&\\
 2 & $A=0, \ d\neq1, \ S=1 $  & $P_t, \  P_x, \  I, \ D=2t\p_t+x\p_x+2(1+t)u\p_u+2tv\p_v$ \\
 \hline &&\\
 3 & $A=0,  \ d=1, \ S\neq1, \ R\neq S$  & $P_t, \  P_x, \  I, \ G=2t\p_x-x\,I$ \\
  \hline &&\\
   4 & $A=0, \ d=1, $  & $P_t, \  P_x, \  I, \ G, $
${\cal{Q}}= e^{(S-1)\,t}\Big(S\,I+(1-S)u\p_v\Big)$
 \\
 &   $R=S\neq1$    &\\
   \hline &&\\
   5 & $A=0, \ d=1, \ R=S=1$  & $P_t, \  P_x, \  I, \ G, \ D, \ Y= t\,I-u\p_v$ \\
  \hline &&\\
   6 & $A=0, \ d=1, \ S=1, R\not=1$  & $P_t, \  P_x, \  I, \ G, \ D,$ \\  &&
   $\Pi=t^2\p_t+tx\p_x+\lf(t^2+\frac{R+1}{2(R-1)}\,t-\frac{x^2}{4}\rg)\,I+$ \\ &&
   $\frac{1}{1-R}\,u\p_v-2tv\p_v$ \\
   \hline
 \end{tabular}
 \end{table}

\textbf{Proof}  of Theorem~\ref{th-1} is based on the classical Lie
method, which is described in detail in several textbooks and
monographs
 (the most recent are \cite{arrigo-15,bl-anco-10,ch-se-pl-book}).
 Applying the invariance criteria  to the DHT system (\ref{1-1}), the system
 of determining equations was derived. Solving the system for arbitrary given parameters $A, \ R, \ S$ and $d$, one immediately obtains
 the trivial  algebra (\ref{2-1}). However, setting additional restrictions on the above
 parameters (see the 2nd column in Table~\ref{tab1}), six extensions of the trivial algebra  can be derived
  (see the 3rd column in Table~\ref{tab1}).

 It should be stressed that all the results presented in Table~\ref{tab1} can be also identified  from the Lie symmetry
 classification of the general two-component system of reaction-diffusion equations derived in \cite{ch-king1,ch-king2}
 provided we restrict ourselves on the 1D space (the above papers deal with systems in  $n$-dimensional space).

 In fact, one easily notes that Cases 1, 3 and 4 of the above table represent particular cases
 of Case 1 of Table 4 \cite{ch-king1}, Case 1 of Table 1 \cite{ch-king1}  and Case 4 of Table 1 \cite{ch-king2}, respectively. Other three cases can be identified  by using additionally the transformation
 \be\label{2-2}
 u \to e^t u,  \quad v \to e^t v.  \ee
 In fact, (\ref{2-2})  reduces  the DHT system (\ref{1-1}) with $A=0$ and $S=1$  to the form
 \be\label{2-3}\ba{l}
u_t = u_{xx}-R\,v, \medskip\\
v_t = dv_{xx}-\frac{v^2}{u}. \ea\ee Now one easily notes that the
reaction-diffusion system (\ref{2-3}) with $d\not=0$ admits the four
dimensional Lie algebra with the basic operators given in Case 3 of
Table 3 \cite{ch-king1} by setting $\alpha_0=\alpha_1=1$ therein.
So, Case 2 of Table~\ref{tab1} is also identified. It can be also
calculated that the Lie symmetry $D$ arising above in
Table~\ref{tab1} simplifies to the form $D_1=2t\p_t+x\p_x+2u\p_u$
via  transformation (\ref{2-2}).

Finally, system (\ref{2-3}) with $d=1$ depending on $R$ admits two
further extensions of Lie symmetry that
 are presented in Cases 1 and 2 of Table 1 \cite{ch-king2}. So, Cases 5 and 6 of Table~\ref{tab1} are  identified.
  Notably, using the above  transformation the Lie symmetry $\Pi$  is reducible to the form
\[\Pi_1=t^2\p_t+tx\p_x+\lf(\frac{R+1}{2(R-1)}\,t-\frac{x^2}{4}\rg)I -2tv\p_v + \frac{1}{1-R}u\p_v,\]
 while the operator $Y$ does not change the form.

 All Lie symmetries presented in Table~\ref{tab1} can be applied for search for exact solutions of  the  DHT system with the relevant restrictions on the parameters
 $A, \ R, \ S$ and $d$. It is worth to note that the reaction-diffusion system (\ref{2-3}) with $d=1$ and its highly nontrivial Lie symmetry was firstly identified in \cite{ch-88} (see also \cite{fu-ch-95}). Some exact solutions were constructed in  \cite{ch-88} as well.

 Here we  examine the  DHT system (\ref{1-1}) with the parameters corresponding to Cases~1 and 3 of Table~\ref{tab1}, i.e.
\be\label{2-11}\ba{l}
u_t = u_{xx}+u-R\,v, \medskip\\
v_t = dv_{xx}+ Sv\lf(1-\frac{v}{u}\rg), \  S\not=1. \ea\ee These
cases are the most interesting from the applicability point of view.
In particular, the linear parts of  the reaction
  terms  of (\ref{2-11}) are not removable by (\ref{2-2}).


 In order to reduce system (\ref{2-11}) to ODE systems, one needs firstly to construct a set of ans\"atze
  corresponding to Lie symmetries.
 There are two general approaches to implement this step.
 The first one consists in examination of the most general linear combination
 of all basic operators  of the Lie algebra of invariance, while  the second  is based on
 systems of inequivalent (nonconjugated) subalgebras that are called optimal systems (see
 more details about advantages/disadvantages of these approaches in  Section\,1.3 of \cite{ch-se-pl-book}).
Because the   Lie algebras of invariance  of system (\ref{2-11}) are
of low dimensionality, we use their optimal systems  of
one-dimensional sub-algebras derived in the seminal work
\cite{Pat-Wint-77}. The optimal systems have the form  (see algebras
$3A_1$ and $A_{4,1}$ in Tables 1 and 2 \cite{Pat-Wint-77},
respectively)\,:
\[P_t+\alpha P_x+\beta I, \ P_x+\beta I, \ I\]
and
\[P_x, \ P_t+\beta I, \ \alpha P_t+G, \ I\]  in  Cases 1 and 3, respectively.
 Here $\alpha$ and $\beta$ are arbitrary constants.

Obviously, the Lie symmetry $I$ is useless for finding exact
solution. One can only claim, using the Lie group  generated by $I$,
that an arbitrary
  solution $(u_0, v_0)$ of (\ref{2-11})  multiplied by an arbitrary constant $C$ produces  another solution.  Application of  the Lie symmetry $P_x$  allows us to look for  space-independent solutions of system (\ref{2-11}), in which we are not interested. Moreover, the operator $P_t+\beta I$ is simply a particular case of that $P_t+\alpha P_x+\beta I$.

Thus,   the Lie symmetries  $P_t+\alpha P_x+\beta I$,  $P_x+\beta I$
and  $\alpha P_t+G$ (in the case $d=1$) produce all inequivalent
ans\"atze for reduction of system (\ref{2-11}) to ODE  systems.
Applying the well-known procedure to the above operators of Lie
symmetry, which are described in any textbook on Lie symmetry
analysis,  the relevant ans\"atze and reduced systems of ODEs were
constructed.   The results are presented in Table~\ref{tab2}.

\begin{table}[h!]
\caption{Ans\"atze and reduced systems of ODEs corresponding to the
DHT system (\ref{2-11})}\medskip
\label{tab2}       
\begin{tabular}{|c|c|c|c|}
\hline  & Operators &  Ans\"atze & Systems of ODEs \\  \hline &&& \\
 1 & $P_t+\alpha P_x+\beta I$  & $u=\varphi(\omega)e^{\beta t},$
$ \omega=x-\alpha\,t$ & $\varphi''+\alpha\varphi'+(1-\beta)\varphi-R\,\psi=0$ \\
   &&$v=\psi(\omega)e^{\beta t}$&
   $d\psi''+\alpha\psi'+(S-\beta)\psi-\frac{S\psi^2}{\varphi}=0$ \\
   \hline &&&\\
 2 & $P_x+\beta I$  & $u=\varphi(t)e^{\beta x}$ & $\varphi'=(1+\beta^2)\varphi-R\,\psi$ \\
   &&$v=\psi(t)e^{\beta x}$&
   $d\psi'=(S+d\beta^2)\psi-\frac{S\psi^2}{\varphi}$ \\
   \hline &&&\\
 3 & $\alpha P_t+G, \ \alpha\neq0$  & $u=\varphi(y)\exp\lf(\frac{2t^3}{3\alpha^2}-\frac{tx}{\alpha}\rg)$ &
 $\alpha^4\varphi''+(\alpha^2-y)\varphi-R\alpha^2\,\psi=0$ \\
   &&$v=\psi(y)\exp\lf(\frac{2t^3}{3\alpha^2}-\frac{tx}{\alpha}\rg)$&
   $\alpha^4\psi''+(S\alpha^2-y)\psi-\frac{S\alpha^2\psi^2}{\varphi}=0$ \\
  & & $y=t^2-\alpha x$ & \\
   \hline &&&\\
 4 & $G$  & $u=\varphi(t)\exp\lf(-\frac{x^2}{4t}\rg)$ & $\varphi'=\lf(1-\frac{1}{2t}\rg)\varphi-R\,\psi$ \\
   &&$v=\psi(t)\exp\lf(-\frac{x^2}{4t}\rg)$&
   $\psi'=\lf(S-\frac{1}{2t}\rg)\psi-\frac{S\psi^2}{\varphi}$ \\
\hline
 \end{tabular}
 \end{table}
 \medskip


 \section { Lie  solutions   of the DHT model} \label{sec-3}

Here we examine the ODE systems arising  in Table~\ref{tab2} in
order to  find exact solutions. Because Lie symmetries are used for
constructing the  ans\"atze and  the reduced systems listed in the
table, the relevant exact solutions  are often called Lie solutions
(another terminology is `group-invariant solutions') of PDEs
(systems of PDEs) in question.

\textbf{Case 1 of Table~\ref{tab2}}. The ODE system
\be\label{2-13}\ba{l}
\varphi''+\alpha\varphi'+(1-\beta)\varphi-R\,\psi=0,\\
d\psi''+\alpha\psi'+(S-\beta)\psi-\frac{S\psi^2}{\varphi}=0\ea\ee
 is reducible to the fourth-order nonlinear ODE
\be\label{2-15}\ba{l}
d\,R\,\varphi^{(4)}+\alpha(1+d)R\,\varphi^{(3)}-S\frac{{\varphi''}^2}{\varphi}-2\alpha
S\frac{\varphi'\varphi''}{\varphi}-\alpha^2S\frac{{\varphi'}^2}{\varphi}+
\medskip \\
\Big(2S(\beta-1)+R\lf(\alpha^2+d+S-\beta-d\beta\rg)\Big)\varphi''+
\alpha\Big(2S(\beta-1)+R\lf(1+S-2\beta\rg)\Big)\varphi'+\medskip\\(1-\beta)\Big(S\lf(\beta-1\rg)+R\lf(S-\beta\rg)\Big)\varphi=0,\ea\ee
because the first equation of the system  can be rewrited as
\[\psi=\frac{1}{R}\big(\varphi''+\alpha\varphi'+(1-\beta)\varphi\big).\]

To the best of our knowledge,  the general solution of the nonlinear
ODE (\ref{2-15}) cannot be found, so that we look for its particular
solutions. Notably the case $\beta=0$ leading  to plane wave
solution of   the  DHT system (\ref{2-11}) is not special if one
wants to solve the above ODEs.

It can be noted that  ODE (\ref{2-15}) simplifies essentially if one
sets $\alpha=0$, i.e. $\omega=x$\,: \[\ba{l}
d\,R\,\varphi^{(4)}-S\frac{{\varphi''}^2}{\varphi}+\Big(2S(\beta-1)+R\lf(d+S-\beta-d\beta\rg)\Big)\varphi''+\\
(1-\beta)\Big(S\lf(\beta-1\rg)+R\lf(S-\beta\rg)\Big)\varphi=0.\ea\]
The latter is reducible to the form \be\label{2-15**}
\varphi\varphi^{(4)}={\varphi''}^2 \ee provided the restrictions
\[S=dR, \ \beta=\frac{d(R-1)}{1-d} \] hold.  Because
the general solution of the nonlinear ODE (\ref{2-15**})  can not be
derived  we look for  its particular solutions in the form
\[\varphi= x^{\gamma}.\] The simple calculations
lead to $\gamma=1$ and $\gamma=\frac{3}{2}$.

In the case $\gamma=\frac{3}{2}$ the
 exact  solution of the DHT system~(\ref{2-11}) with $S=dR$ is constructed  in the form
\[\ba{l}
u(t,x)=x^{\frac{3}{2}}\exp\lf(\frac{d(R-1)}{1-d}\,t\rg),\medskip\\
v(t,x)=\lf(\frac{3}{4R}\,x^{-\frac{1}{2}}+\frac{1-dR}{(1-d)R}\,x^{\frac{3}{2}}\rg)\exp\lf(\frac{d(R-1)}{1-d}\,t\rg).
\ea\] Other particular solutions of  ODE (\ref{2-15**}) are
presented below (see formula (\ref{2-36})).

Now we  return to the ODE system (\ref{2-13}) and use
the assumption about  linear dependenance of  the functions
$\varphi$ and $\psi$.
 Having the above assumption, one easily integrates  the first equation of system
(\ref{2-13}) because it is the linear ODE with constant
coefficients. Substituting the obtained functions $\varphi$ and
$\psi$ into the second equation of system (\ref{2-13}), one arrives
at an algebraic equation. Two different cases, $d=1$  and $d\not=1$,
should be examined. So,
 making the
relevant calculations we obtain  the  results  presented below.\\

\emph{1. If $d=1$ then} \[\ba{l}\varphi(\omega)=\left\{
\begin{array}{l} C_1\exp\lf(-\frac{2\kappa+\alpha}{2}\,\omega\rg)+
C_2\exp\lf(\frac{2\kappa-\alpha}{2}\,\omega\rg), \quad
   \mbox{if} \quad \beta>\frac{(R-1)S}{R-S}-\frac{\alpha^2}{4},
 \medskip \\
 C_1\sin(\kappa\,\omega+C_0)\exp\lf(-\frac{\alpha}{2}\,\omega\rg),
 \quad \mbox{if} \quad  \beta<\frac{(R-1)S}{R-S}-\frac{\alpha^2}{4},
 \medskip  \\
 \big(C_1+C_2\,\omega\big)\exp\lf(-\frac{\alpha}{2}\,\omega\rg), \quad \mbox{if} \quad \beta=\frac{(R-1)S}{R-S}-\frac{\alpha^2}{4},
\end{array} \right. \medskip \\
\psi(\omega)=\frac{1-S}{R-S}\,\varphi(\omega), \ea\] where
$\kappa=\sqrt{\lf|\beta+\frac{\alpha^2}{4}-\frac{(R-1)S}{R-S}\rg|}$\,.
Hereafter $C$ with and without subscripts are arbitrary constants.

Substituting the functions $\varphi$ and $\psi$ into the ansatz from
Case~1 of Table~\ref{tab2}, we obtain three different  solutions of
the DHT system~(\ref{2-11}) with $d=1$:
\[\ba{l}u(t,x)=\left\{
\begin{array}{l}
C_1\exp\lf(-\lf(\kappa+\frac{\alpha}{2}\rg)\,x+\frac{2\beta+\alpha^2+2\alpha\,\kappa}{2}\,t\rg)+\\
\hskip2cm
C_2\exp\lf(-\lf(\kappa-\frac{\alpha}{2}\rg)\,x+\frac{2\beta-\alpha^2+2\alpha\,\kappa}{2}\,t\rg),
\
   \mbox{if} \ \beta>\frac{(R-1)S}{R-S}-\frac{\alpha^2}{4},
 \medskip \\ \sin\big(\kappa\,\lf(x-\alpha t\rg)+C_0\big)
 \exp\lf(-\frac{\alpha}{2}\,x+\lf(\beta+\frac{\alpha^2}{2}\rg)\,t\rg),
 \ \mbox{if} \ \beta<\frac{(R-1)S}{R-S}-\frac{\alpha^2}{4},
 \medskip  \\
 \big(C_1+C_2\,(x-\alpha t)\big) \exp\lf(-\frac{\alpha}{2}\,x+\lf(\frac{\alpha^2}{4}+\frac{(R-1)S}{R-S}\rg)\,t\rg),
  \ \mbox{if} \  \beta=\frac{(R-1)S}{R-S}-\frac{\alpha^2}{4},
\end{array} \right. \medskip \\
v(t,x)=\frac{1-S}{R-S}\,u(t,x), \ea\] where
$\kappa=\sqrt{\lf|\beta+\frac{\alpha^2}{4}-\frac{(R-1)S}{R-S}\rg|}$.

\emph{2. If  $d\not=1$ then} one may set  $\alpha=0$ (the choice
$\alpha\not=0$ leads to the same solutions of the DHT
system~(\ref{2-11})), so that $\omega=x$. In this case
\be\label{2-36}\ba{l}\varphi(x)=\left\{
\begin{array}{l} C_1e^{-\kappa\,x}+
C_2e^{\kappa\,x}, \quad
   \mbox{if} \quad \frac{R-S}{dR-S}\beta+\frac{1-R}{dR-S}S>0,
 \medskip \\ C_1\sin\lf(\kappa\,x+C_0\rg),
 \quad \mbox{if} \quad  \frac{R-S}{dR-S}\beta+\frac{1-R}{dR-S}S<0,
 \medskip  \\
C_1+C_2\,x, \quad \mbox{if} \quad \beta=\frac{(R-1)S}{R-S},
\end{array} \right. \medskip \\
\psi(x)=\frac{d-S+(1-d)\beta}{dR-S}\,\varphi(x), \ea\ee where
$\kappa=\sqrt{\lf|\frac{R-S}{dR-S}\beta+\frac{1-R}{dR-S}S\rg|}, \
dR-S\neq0$\,.

Thus, using the obtained above  functions $\varphi$ and $\psi$, we
again obtain three different  solutions of the DHT
system~(\ref{2-11}):
 \[\ba{l}u(t,x)=\left\{
\begin{array}{l} C_1e^{-\kappa\,x+\beta t}+
C_2e^{\kappa\,x+\beta t}, \quad
   \mbox{if} \quad \frac{R-S}{dR-S}\beta+\frac{1-R}{dR-S}S>0,
 \medskip \\ \sin\lf(\kappa\,x+C_0\rg)e^{\beta\,t},
 \quad \mbox{if} \quad  \frac{R-S}{dR-S}\beta+\frac{1-R}{dR-S}S<0,
 \medskip  \\
(C_1+C_2\,x)e^{\beta\,t}, \quad \mbox{if} \quad
\beta=\frac{(R-1)S}{R-S},
\end{array} \right. \medskip \\
v(t,x)=\frac{d-S+(1-d)\beta}{dR-S}\,u(t,x), \
\kappa=\sqrt{\lf|\frac{R-S}{dR-S}\beta+\frac{1-R}{dR-S}S\rg|}. \ea\]

\textbf{Case 2 of Table~\ref{tab2}}. The relevant reduced system
consists of first-order ODEs, hence that  is reducible to the
following  second-order ODE
\be\ba{l}\label{2-26}R\varphi''-S\frac{{\varphi'}^2}{\varphi}+\big(2S(1+\beta^2)-R(1+S+(1+d)\beta^2)\big)\varphi'+\medskip\\
\lf(1+\beta^2\rg)\big(R\lf(S+d\beta^2\rg)-S\lf(1+\beta^2\rg)\big)\varphi=0,\ea\ee
while  the function $\psi(t)$ is defined as follows
\[\psi=\frac{1}{R}\big(-\varphi'+\lf(1+\beta^2\rg)\varphi\big).\]
Applying the substitution
\be\label{2-23}\frac{\varphi'}{\varphi}=\chi(t)\ee to the nonlinear
ODE (\ref{2-26}), we obtain  the Riccati equation \[\ba{l}
R\chi'=(S-R)\chi^2-\big(2S(1+\beta^2)-R(1+S+(1+d)\beta^2)\big)\chi-\medskip\\
\lf(1+\beta^2\rg)\big(R\lf(S+d\beta^2\rg)-S\lf(1+\beta^2\rg)\big).
\ea\] The  general solution of the latter is readily constructed and
has  three different forms depending on the parameters $R, \ S$ and
$d$\,: \be\label{2-28}\chi(t)=1+\beta^2-\frac{\gamma
R}{{(R-S)(1+C_1e^{\gamma t})}}, \ \gamma\equiv
1-S+(1-d)\beta^2\neq0, \ R-S\neq0;\ee
\[\chi(t)=1+\beta^2+C_1e^{-\gamma t}, \ \gamma\neq0, \
R=S;\]
\[\chi(t)=1+\beta^2+\frac{R}{{C_1+\lf(R-1+(d-1)\beta^2\rg)t}},\
\gamma=0. \]
So, three different exact solutions  of the DHT system~(\ref{2-11})
can be constructed. In particular, using the formulae  (\ref{2-28})
and (\ref{2-23}) and the  ansatz from Case~2 of Table~\ref{tab2}, we
obtain
  the exact solution \[\ba{l}
u(t,x)=\lf(C+e^{-\gamma t}\rg)^{\frac{R}{R-S}}\exp\big((1+\beta^2)t+\beta x\big), \medskip\\
v(t,x)=\frac{\gamma}{R-S}\,\lf(C+e^{-\gamma
t}\rg)^{\frac{S}{R-S}}\exp\big((S+d\beta^2)t+\beta x\big), \ \ea\]
 where $C$ and $\beta$  are arbitrary constants and  $ \gamma=1-S+(1-d)\beta^2$.

\textbf{Case 3 of Table~\ref{tab2}}.   The relevant reduced system
\be\label{2-30}\ba{l}
\alpha^4\varphi''+(\alpha^2-y)\varphi-R\alpha^2\,\psi=0, \quad \alpha\not=0, \\
\alpha^4\psi''+(S\alpha^2-y)\psi-\frac{S\alpha^2\psi^2}{\varphi}=0
\ea\ee is similar to that arising in Case 1 of Table~\ref{tab2}  but
involves two additional linear terms.  Similarly to the ODE
system~(\ref{2-13}), we do not expect that  it is possible to
construct the general solution of (\ref{2-30}). However,  particular
solutions can identified by applying additional restrictions. For
example, assuming a linear dependence between the functions
$\varphi$ and $\psi$,  we arrive at the linear second-order ODE for
$\varphi$ \be\label{2-31} \varphi''
-\lf(\alpha^{-4}y+\alpha^{-2}b\rg)\varphi=0, \quad
b=\frac{S(1-R)}{(R-S)},\ee while the function
$\psi(y)=\frac{S-1}{S-R}\,\varphi(y)$ (here $S\neq R$). It is
well-known that ODE (\ref{2-31}) is reducible to the Airy equation $
\varphi''(z)=z\varphi(z)$ by the substitution
$z=\alpha^{-4/3}y+\alpha^{2/3}b$. Thus, the  general solution of ODE
(\ref{2-31}) is \[ \varphi(y)=C_1
Ai\lf(\alpha^{-\frac{4}{3}}y+\frac{\alpha^{\frac{2}{3}}(1-R)S}{R-S}\rg)+
C_2
Bi\lf(\alpha^{-\frac{4}{3}}y+\frac{\alpha^{\frac{2}{3}}(1-R)S}{R-S}\rg),\]
where $Ai$  and $Bi$ are the Airy functions of the first and second
kind, respectively. These functions can be expressed via the
modified Bessel functions $I_{1/3}$ and $I_{-1/3}$ using the
well-known formulae.

Now we  substitute  the  obtained functions $\varphi$ and $\psi$
into the ansatz from Case~3 of Table~\ref{tab2}  and obtain
three-parameter family of exact solutions  \[\ba{l} u(t,x)=\lf(C_1
Ai\lf(\alpha^{-\frac{4}{3}}(t^2-\alpha
x)+\frac{\alpha^{\frac{2}{3}}(1-R)S}{R-S}\rg)+ \rg.
\\ \lf. \hskip3cm
C_2 Bi\lf(\alpha^{-\frac{4}{3}}(t^2-\alpha
x)+\frac{\alpha^{\frac{2}{3}}(1-R)S}{R-S}\rg)\rg)\exp\lf(\frac{2t^3}{3\alpha^2}-
\frac{tx}{\alpha}\rg),\medskip \\
 v(t,x)=\frac{S-1}{S-R}\,u(t,x)\ea\] of the DHT system~(\ref{2-11})
with $d=1$.

Finally, we examine \textbf{Case~4 of Table~\ref{tab2}}.
In this case, the ODE system has a similar structure to that in
Case~2. So, the same approach for its solving is applicable. The
relevant second-order ODE has the form \be\ba{l}\label{2-20}
R\varphi''-S\frac{{\varphi'}^2}{\varphi}+\lf(2S-R-RS+\frac{R-S}{t}\rg)\varphi'+\medskip\\
\lf(S(R-1)-\frac{R+S}{4t^2}+\frac{2S-R-RS}{2t}\rg)\varphi=0\ea\ee
and the function
\[\psi=\frac{1}{R}\Big(-\varphi'+\lf(1-\frac{1}{2t}\rg)\varphi\Big). \]
Applying the substitution (\ref{2-23}) to ODE (\ref{2-20}), one
again obtains  the Riccati equation \be\label{2-21}
R\chi'=(S-R)\chi^2+\lf(\frac{S-R}{t}+R+RS-2S\rg)\chi+\frac{R+S}{4t^2}+\frac{R+RS-2S}{2t}+S\lf(1-R\rg).
\ee   It can be noted  that the function $\chi_p=1-\frac{1}{2t}$ is
a particular solution of ODE (\ref{2-21}). So, the  general solution
of the Riccati equation (\ref{2-21}) is readily constructed
\be\label{2-22}\chi(t)=1-\frac{1}{2t}+\frac{R\lf(S-1\rg)}{R-S}\frac{1}{1+Ce^{\lf(1-S\rg)t}},
\ee\ provided  $R\neq S.$
If $R= S$  then the  general solution is \[
\chi(t)=1-\frac{1}{2t}+Ce^{\lf(S-1\rg)t}. \]

Thus, using the formulae  (\ref{2-22}) and (\ref{2-23}) and the
ansatz from Case~4 of Table~\ref{tab2}, we obtain
  the following exact solution  of the DHT system~(\ref{2-11})
with $d=1$\,:
   \be\label{2-24}\ba{l}
u(t,x)=\frac{\exp\lf(t-\frac{x^2}{4t}\rg)}{\sqrt{t}}\lf(C+e^{\lf(S-1\rg)t}\rg)^{\frac{R}{R-S}}, \medskip\\
v(t,x)=\frac{S-1}{S-R}\,\frac{\exp\lf(St-\frac{x^2}{4t}\rg)}{\sqrt{t}}\lf(C+e^{\lf(S-1\rg)t}\rg)^{\frac{S}{R-S}}.\ea\ee
If an arbitrary constant is specified as   $C=1$ then the exact
solution (\ref{2-24}) takes the form \be\label{2-34}\ba{l}
u(t,x)=\frac{1}{\sqrt{t}}\cosh\lf(\frac{S-1}{2}\,t\rg)^{\frac{R}{R-S}}\exp\lf(\frac{2S-R-RS}{2(S-R)}\,t-\frac{x^2}{4t}\rg),
 \ R\neq S, \ S\neq1, \medskip\\
v(t,x)=\frac{S-1}{2(S-R)}\lf(1+\tanh\lf(\frac{S-1}{2}\,t\rg)\rg)u(t,x).\ea\ee

Since DHT system~(\ref{2-11}) admits the time translation, the exact
solution (\ref{2-34}) can be rewritten in the form
\be\label{2-34*}\ba{l}
u(t,x)=\frac{1}{\sqrt{t+t_0}}\cosh\big(\frac{S-1}{2}\,(t+t_0)\big)^{\frac{R}{R-S}}
\exp\lf(\frac{2S-R-RS}{2(S-R)}\,t-\frac{x^2}{4(t+t_0)}\rg),
  \medskip\\
v(t,x)=\frac{S-1}{2(S-R)}\Big[1+\tanh\lf(\frac{S-1}{2}\,(t+t_0)\rg)\Big]u(t,x),
 \ R\neq S, \ S\neq1,\ea\ee where $t_0>0$ is an arbitrary parameter.

\begin{figure}[h!]
\begin{center}
\includegraphics[width=6cm]{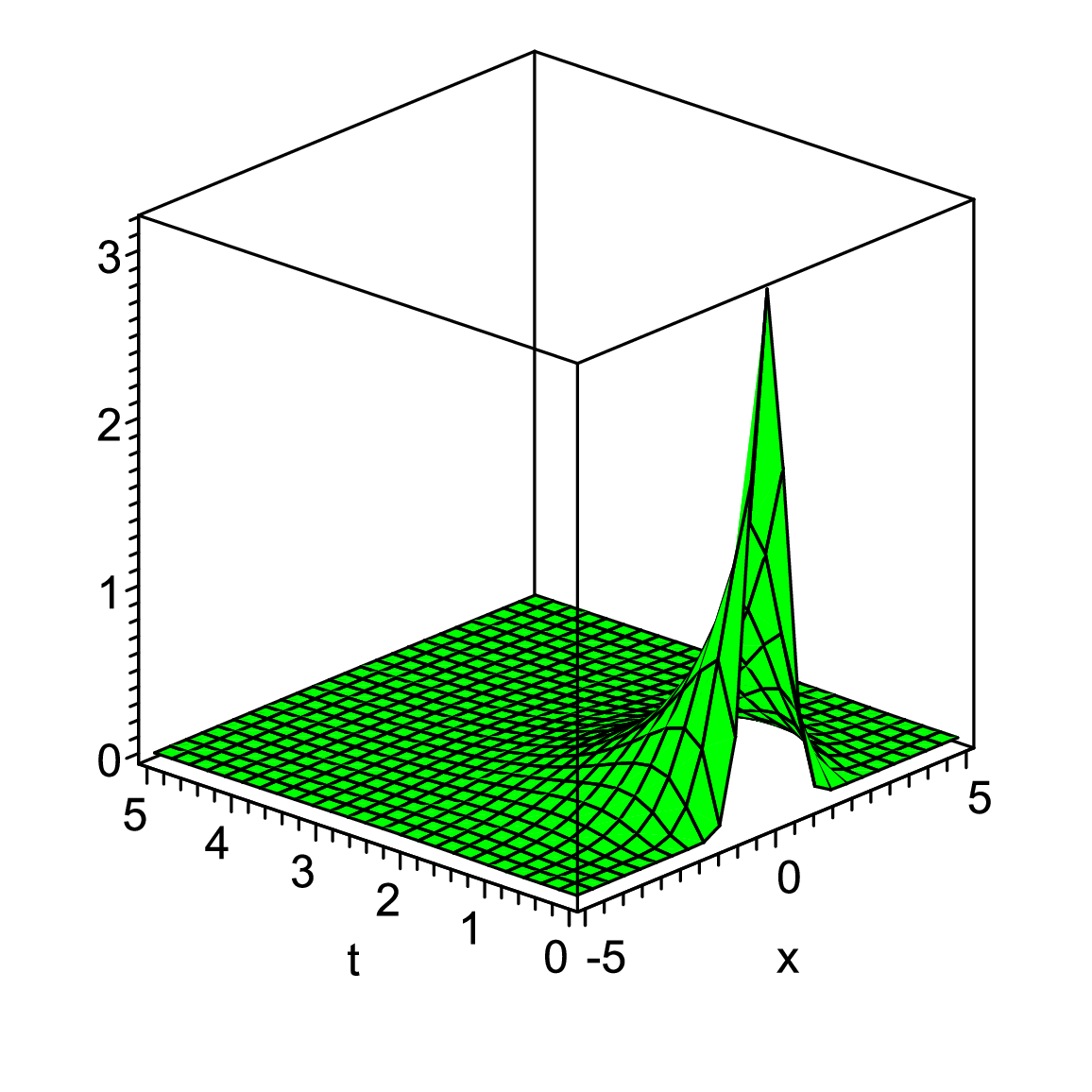}
\includegraphics[width=6cm]{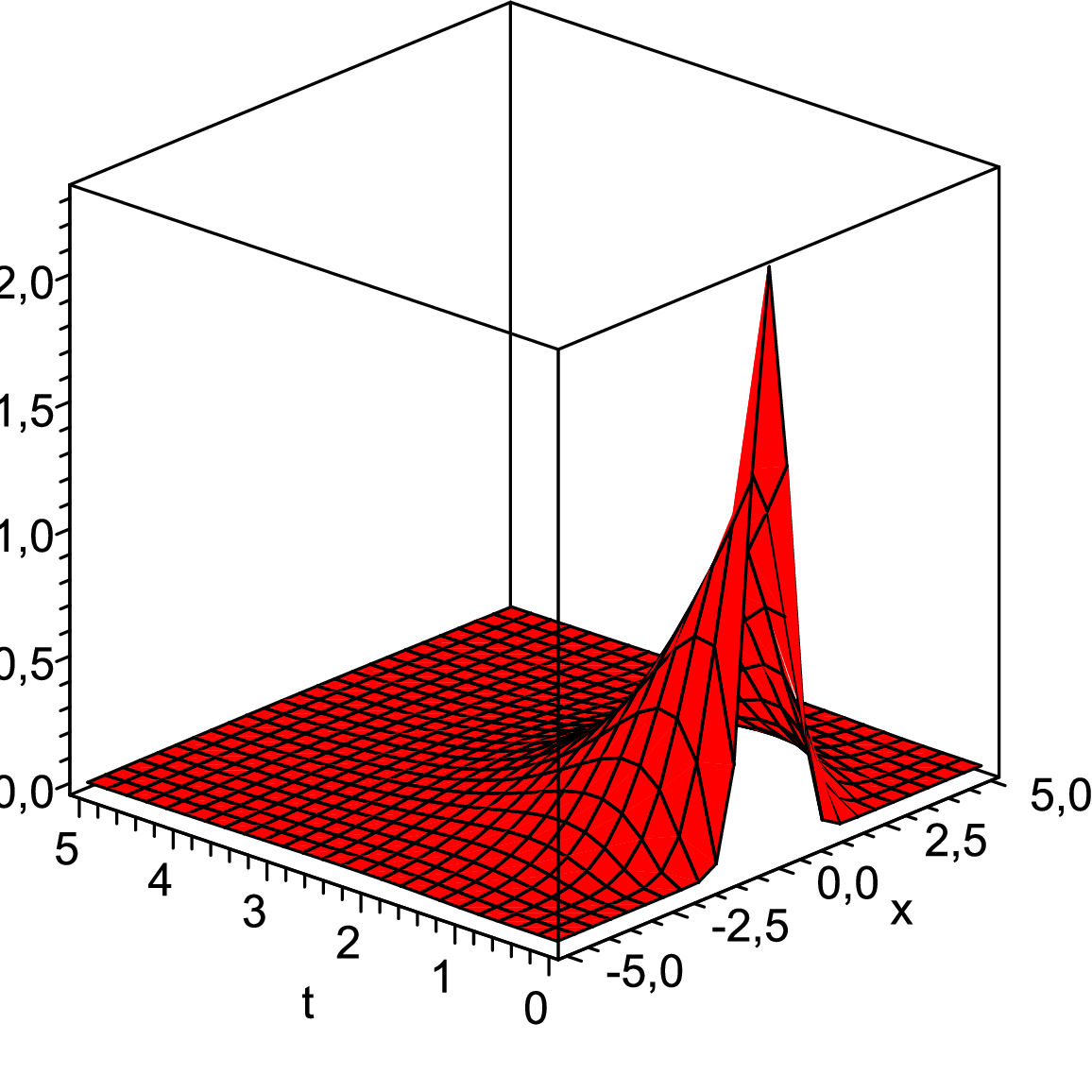}
\end{center}
\caption{Surfaces representing the $u$ (green) and $v$ (red)
components  of solution (\ref{2-34*})
 of the DHT system (\ref{2-11}) with the parameters $d=1, \ S=3, \ R=1.5, \ t_0=0.1.$} \label{f1}
\end{figure}

\begin{figure}[h!]
\begin{center}
\includegraphics[width=6cm]{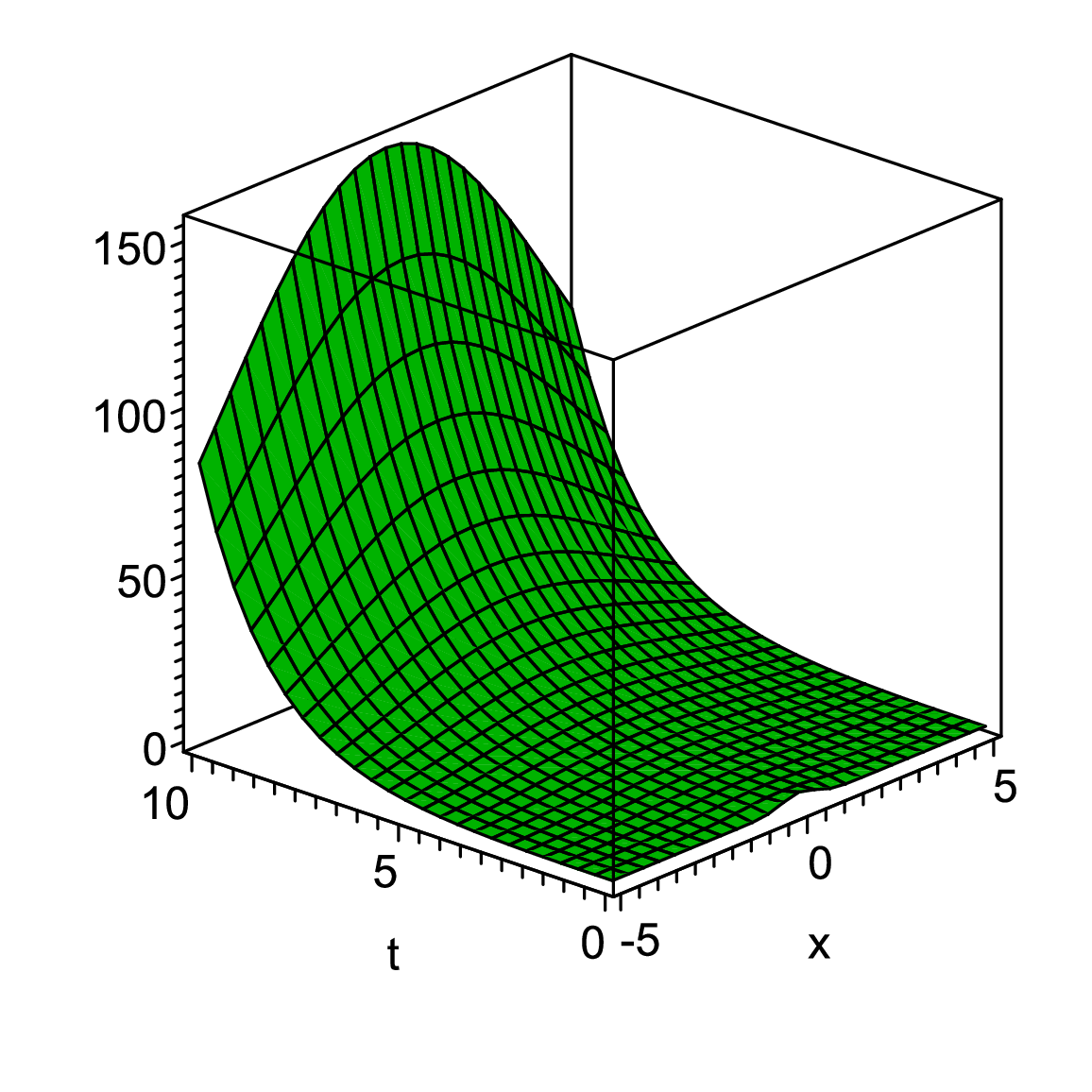}
\includegraphics[width=6cm]{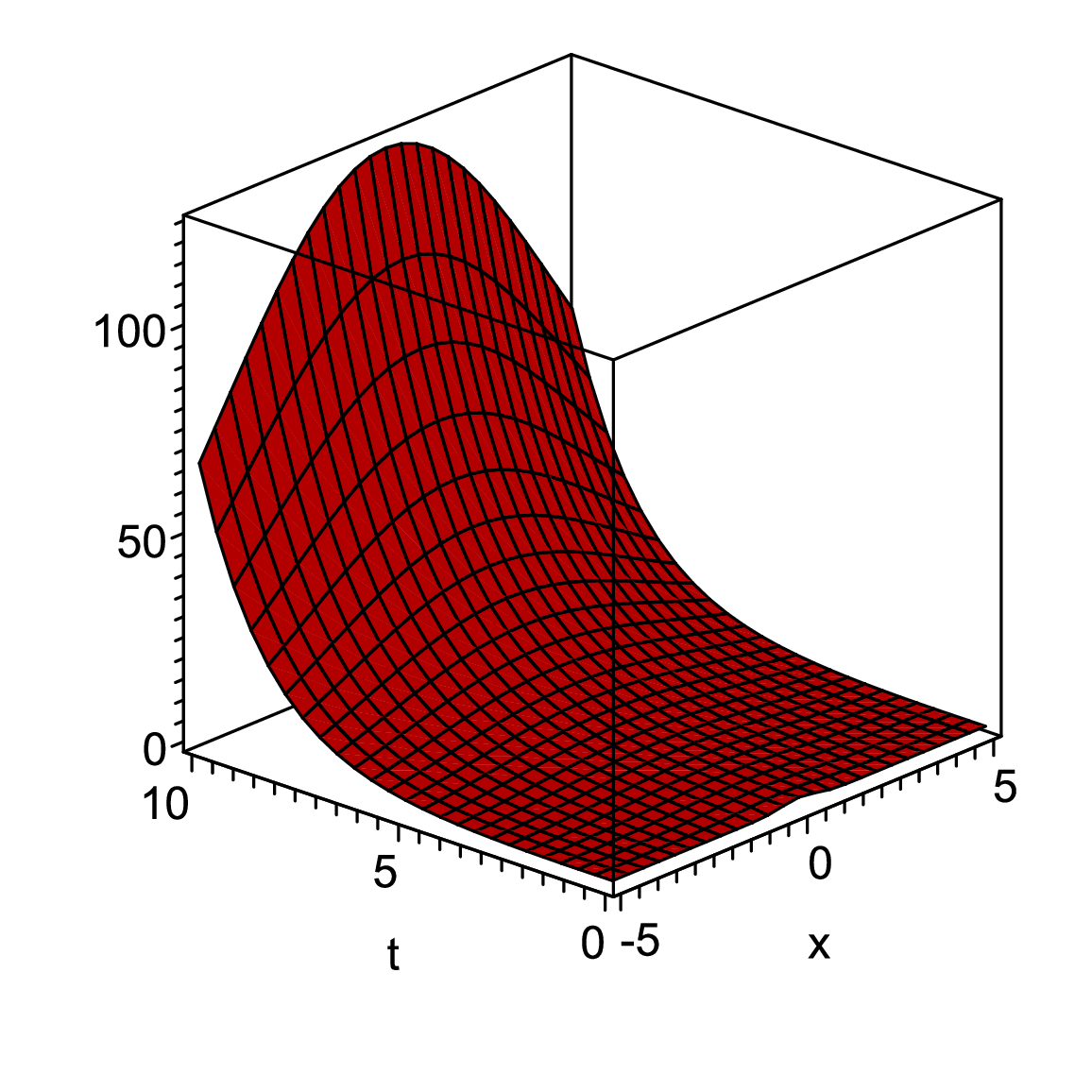}
\end{center}
\caption{Surfaces representing the $u$ (green) and $v$ (red)
components  of solution (\ref{2-34*})
 of the DHT system (\ref{2-11}) with the parameters $d=1, \ S=3, \ R=0.5, \ t_0=0.1.$} \label{f2}
\end{figure}

 It should be noted that  both  components of solution (\ref{2-34*}) are
  bounded and nonnegative in the domain \[\Omega=\left\{ (t,x) \in (0,+ \infty )\times
(-\infty,+\infty)\rg\}\] provided the coefficient restrictions
\[S>R\geq1\] hold.
Examples of solution (\ref{2-34*}) are presented in Figures~\ref{f1}
and~\ref{f2}. If we turn to the biological sense of this exact
 solution then one notes that the plots in Fig.\,\ref{f1}  demonstrate
 the complete extinction of the prey $u$ and the predators $v$ as time
 $t \to +\infty$. The plots in Fig.\,\ref{f2} represent unbounded growth of both species $u$ and $v$. It may happen because we assumed that the ecological system produces unbounded amount of food for preys (see Introduction), i.e.  the Malthusian law is assumed for the prey growth.

In conclusion of this section, we highlight a nontrivial formula for
the  solution multiplication by using the Lie symmetry of  the  DHT
system (\ref{2-11}). In fact, this system with $d=1$ admits
four-dimensional Lie algebra (see Case 3 in Table \ref{tab1}), which
is nothing else but the Galilei algebra in 1D space $AG(1.1)$
\cite{fu-ch-95}. Thus, applying continuous transformations from the
four-parameter Lie group that   corresponds  to $AG(1.1)$  to a
given exact solution $(u_0(t,x),v_0(t,x))$,  one obtains the new
solution \be\label{2-25} \ba{l}
\medskip
 u_{new}=Cu_0
(t+t_0, x + \varepsilon t+x_0) \exp \left(\frac{\varepsilon}{2}
\left( x+\frac{\varepsilon t}{2}\right) \right), \\
 v_{new}=Cv_0
(t+t_0, x + \varepsilon t+x_0) \exp \left(\frac{\varepsilon}{2}
\left( x+\frac{\varepsilon t}{2}\right) \right) \ea \ee of system
(\ref{2-11}) with $d=1$. Here $C,\ t_0, \ x_0$ and $\varepsilon$ are
arbitrary parameters, hence (\ref{2-25}) represents a four-parameter
family of exact solutions. This formula is well-known and  valid for
any Galilei-invariant system of PDEs (see, e.g.,
\cite{ch-88,ch-king1}). Although the solution (\ref{2-25}) is
equivalent to $(u_0(t,x),v_0(t,x))$ according to the classification
of group-invariant solutions of PDEs \cite{olv-93} (see Chapter 3),
such type formulae are helpful from the applicability point of view.
In particular, the above formula allows us  to generate
time-dependent solutions from stationary.

\section{Conditional symmetries of the DHT model (\ref{1-1}) } \label{sec-4}

Let us consider the general form of $Q$-conditional (nonclassical)
symmetry operator of system~(\ref{1-1})  \be \label{3-2}
 Q=\xi^0(t,x,u,v)\partial_t+\xi^1(t,x,u,v)\partial_x+\eta^1(t,x,u,v)\partial_u+\eta^2(t,x,u,v)\partial_v, \ \lf(\xi^0\rg)^2+\lf(\xi^1\rg)^2\neq0,
\ee where    $\xi^i(t,x,u,v)$ and $\eta^k(t,x,u,v)$ are
to-be-determined smooth functions. Because  the DHT system
(\ref{1-1}) is a system of evolution equations, the problem of
constructing its $Q$-conditional symmetry operators of the form
(\ref{3-2})
 depends on the value of the function $\xi^0$  and  one
needs to consider two essentially  different cases\,:
\begin{enumerate}
  \item $\xi^0\neq0 \Rightarrow \xi^0=1.$
  \item $\xi^0=0, \ \xi^1\neq0. $
\end{enumerate}

 Typically, the first case is under study and it is widely thought that the most interesting conditional symmetries are those with $\xi^0\neq0$.
 The algorithm for finding $Q$-conditional symmetries is  well-known and one is based  the  following
 criterion (see, e.g., Chapter 5 in \cite{bl-anco-10}).
\begin{definition} \label{d1}
Operator (\ref{3-2}) is called the $Q$-conditional symmetry for the
HT system (\ref{1-1}) if  the following invariance conditions are
satisfied: \[\ba{l}
\mbox{\raisebox{-1.6ex}{$\stackrel{\displaystyle  
Q}{\scriptstyle 2}$}}  
\lf(u_{xx}-u_t+ u\lf(1-\frac{R\,v}{u+A}\rg)\rg)  
\Big\vert_{{\cal{M}}}=0, \\[0.3cm]  
\mbox{\raisebox{-1.6ex}{$\stackrel{\displaystyle  
Q}{\scriptstyle 2}$}}  
\lf(d v_{xx}-v_t+Sv\lf(1-\frac{v}{u}\rg)\rg)  
\Big\vert_{{\cal{M}}}=0, \ea \]
 where the manifold
 \[\ba{l}{\cal{M}}=
\Big\{S_1=0,S_2=0,Q(u)=0,Q(v)=0,\medskip \\ \frac{\partial}{\partial
t}\,Q(u)=0,
 \frac{\partial}{\partial x}\,Q(u)=0,\frac{\partial}{\partial t}\,Q(v)=0,
  \frac{\partial}{\partial x}\,Q(v)=0\Big\},\ea\] while
 \be\nonumber\ba{l}
 S_1 \equiv \  u_{xx}-u_t+ u\lf(1-\frac{R\,v}{u+A}\rg),\\
S_2 \equiv \   d v_{xx}-v_t+Sv\lf(1-\frac{v}{u}\rg).\ea\ee
\end{definition}
Hereafter we use the notations \[\ba{l} Q(u)=\xi^0u_t+\xi^1u_x-\eta^1, \ Q(v)=\xi^0v_t+\xi^1v_x-\eta^2, \medskip \\
\mbox{\raisebox{-1.7ex}{$\stackrel{\displaystyle Q}{\scriptstyle
2}$}}=Q+\rho^1_t\p_{u_{t}}+\rho^1_x\p_{u_{x}}+\rho^2_t\p_{v_{t}}+\rho^2_x\p_{v_{x}}+\\
\hskip2cm \sigma^1_{tt}\p_{u_{tt}}
+\sigma^1_{tx}\p_{u_{tx}}+\sigma^1_{xx}\p_{u_{xx}}+\sigma^2_{tt}\p_{v_{tt}}
+\sigma^2_{tx}\p_{v_{tx}}+\sigma^2_{xx}\p_{v_{xx}},\ea\] where the
coefficients $\rho$ and $\sigma$ with relevant subscripts are
expressed via the functions $\xi^i$ and $\eta^k$ by the well-known
formulae. As it is shown in \cite{ch-dav-book}, the manifold
${\cal{M}}$ can be replaced by
 \[{\cal{M^*}}=\lf\{S_1=0,S_2=0,Q(u)=0,Q(v)=0,\frac{\partial}{\partial t}\,Q(u)=0, Q(v)=0\rg\}\]
in the case of evolution systems of PDEs. It turns out that
application of the above definition for finding $Q$-conditional
symmetries leads to rather trivial results provided $\xi^0\neq0$.


\begin{theo}\label{th-2} The DHT system (\ref{1-1}) admits only such
$Q$-conditional operators of the form (\ref{3-2}) with $\xi^0\neq0$,
which are equivalent either to the Lie symmetry operators presented
in Table~\ref{tab1} or to their linear combinations. \end{theo}
 In
the case $\xi^0=0$,  operator (\ref{3-2}) takes the form \be
\label{3-4}
 Q=\xi^1(t,x,u,v)\partial_x+\eta^1(t,x,u,v)\partial_u+\eta^2(t,x,u,v)\partial_v, \
 \xi\neq0.
\ee

It is well-known that the task of constructing  the $Q$-conditional
symmetries with $\xi^0=0$ is much more complicated. In fact, the
so-called  determining equations (DEs) obtained by applying
Definition~\ref{d1} are so complicated that can  be solved  only
under additional restrictions. It is generally accepted that
constructing of a general solution of  DEs for a given nonlinear
system  is equivalent to solving of  the system in question  (in the
case of scalar evolution equations this was shown in
\cite{zhdanov}). Very recently, it was noted   the  case  $\xi^0=0$
can be completely examined using the notion of $Q$-conditional
symmetry of the first type \cite{ch-dav-math-21}. Although this
notion was introduced in 2010 \cite{ch-2010},  its  successful
applications were  realized only in the case  $\xi^0\neq0$ (see
\cite{ch-dav-book} and references therein). In the case  $\xi^0=0$,
the invariance criteria for  $Q$-conditional symmetry of the first
type can be formulated as follows.

\begin{definition} \label{d2}
Operator (\ref{3-4}) is called the $Q$-conditional symmetry of the
first type  for the DHT system (\ref{1-1}) if  the following
invariance conditions are satisfied\,: \be\label{3-5}\ba{l}
\mbox{\raisebox{-1.6ex}{$\stackrel{\displaystyle  
Q}{\scriptstyle 2}$}}\, (S_1)
\Big\vert_{{\cal{M}}_1}=0, \\[0.3cm]  
\mbox{\raisebox{-1.6ex}{$\stackrel{\displaystyle  
Q}{\scriptstyle 2}$}}\, (S_2) \Big\vert_{{\cal{M}}_1}=0, \ea \ee
 where the manifold ${\cal{M}}_1$ is either ${\cal{M}}_1^u=\{S_1=0,S_2=0,Q(u)=0,\frac{\partial}{\partial t}\,Q(u)=0, \frac{\partial}{\partial x}\,Q(u)=0\}$ or ${\cal{M}}_1^v=\{S_1=0,S_2=0,Q(v)=0,\frac{\partial}{\partial t}\,Q(v)=0, \frac{\partial}{\partial x}\,Q(v)=0\}$.
\end{definition}

It can be noted that formulae (\ref{3-5}) have the same structure at
those in Definition~\ref{d1}, however, the manifolds ${\cal{M}}_1^u$
and ${\cal{M}}_1^v$ do not coincide with ${\cal{M}}$. It should be
stressed that each $Q$-conditional symmetry of the first type is
automatically a $Q$-conditional (nonclassical) symmetry but not wise
versa \cite{ch-2010}.

 Since system (\ref{1-1}) does not possess  the symmetric structure we
 should apply Definition~\ref{d2}  twice, using both  manifolds ${\cal{M}}_1^u$
 and ${\cal{M}}_1^v$. As a result, the following theorem can be formulated.

\begin{theo}\label{th-3} The DHT system (\ref{1-1}) is invariant under a
$Q$-conditional symmetry of the first type (\ref{3-4})
 if and only if the system and the
corresponding symmetry operator possess the forms listed below.

\textbf{Case I.}  $d\neq 1$\,:\be\label{3-6}\ba{l}
u_t = u_{xx}+u-Sv, \medskip\\
v_t = dv_{xx}+ Sv\lf(1-\frac{v}{u}\rg), \ea\ee
\[Q^u_1=\partial_x+f(t)u\p_u+\lf(f(t)v-\frac{f'(t)}{S}\,u\rg)\p_v,\] where
the smooth function f(t) is the solution of the nonlinear ODE
\be\label{3-7*} f''+(1-d)f^2f'+\lf(1-S\rg)f'=0,\ee which can be
presented in an implicit form \be\label{3-7}
\int\frac{1}{\frac{d-1}{3}f^3+(S-1)f+C_1}df=t+C_2.\ee

 \textbf{Case II.}
$d=1$\,:\be\label{3-8}\ba{l}
u_t = u_{xx}+u-Sv, \medskip\\
v_t = v_{xx}+ Sv\lf(1-\frac{v}{u}\rg), \ea\ee
\[Q^u_2=2g(t)\partial_x+\lf(2h(t)-g'(t)x\rg)u\p_u+\lf(\lf(2h(t)-g'(t)x\rg)v-\frac{\lf(2h'(t)-g''(t)x\rg)}{S}\,u\rg)\p_v,\] where
the smooth functions g(t) and h(t) are the solution of the nonlinear
ODE system \be\ba{l}\label{3-9} hg''+g\lf(h''+\lf(1-S\rg)h'\rg)=0,
\medskip
\\ gg''=Ce^{(S-1)t}.
\ea\ee
 \end{theo}
 \begin{Remark} \label{rem1} In Teorem~\ref{th-3}, the upper index $u$ means that the relevant
 $Q$-conditional symmetry operator satisfy
  Definition~\ref{d2} with  the manifold ${\cal{M}}_1^u$.
 Application of  Definition~\ref{d2} with  the manifold ${\cal{M}}_1^v$ leads only to Lie symmetries.
\end{Remark}

\textbf{Proof} of Theorem 3  consists of the same steps as those
presented in detail in \cite{ch-dav-math-21}  for the diffusive
Lotka--Volterra  system. Thus, the  relevant analysis and
calculations are   omitted here.

\begin{Remark} \label{rem3}
We were unable to construct the general solution
 of the nonlinear ODE system (\ref{3-9}).
However,  the nontrivial  particular  exact solution
\be\label{3-10}g(t)=C_1e^{\frac{S-1}{2}\,t}, \quad
h(t)=\lf(C_2+C_3t\rg)e^{\frac{S-1}{2}\,t}, \ S\neq1\ee was
identified.
\end{Remark}

\begin{Remark} \label{rem3*}
Theorem~\ref{th-3} gives an exhaustive description of
$Q$-conditional symmetries of the first type of the DHT system
(\ref{1-1}). However, one does not present all possible
$Q$-conditional (nonclassical) symmetries because $Q$-conditional
symmetries of the first type form only a subset of nonclassical
symmetries (see \cite{ch-2010} for details).
\end{Remark}

In conclusion of this section, it should be pointed out that the
$Q$-conditional symmetries obtained do not coincide with the Lie
symmetries presented in Table~\ref{tab1}. In fact, the operator
$Q^u_1$ involves a nonconstant function $f(t)$, so that one differs
from the Lie symmetries listed in Cases 1 and 2 of Table~\ref{tab1}.

The operator $Q^u_2$  does not coincide with the Lie symmetries
listed in Cases 4 and 5 of Table~\ref{tab1} provided $g''\neq0$. If
$g(t)$ is  a linear  function then $h''+\lf(1-S\rg)h'=0$. Thus,  the
operator $Q^u_2$ takes  the forms
\[\alpha_1P_x+\alpha_2\,I+\alpha_3\,G+\alpha_4\, {\cal{Q}}
\]
and
\[\alpha_1P_x+\alpha_2\,I+\alpha_3\,G+\alpha_4\,Y\]
in the cases $S\neq1$ and $S=1$, respectively.

\section{Non-Lie exact solutions of the DHT model} \label{sec-5}
Here  our aim is to construct exact solutions of the DHT systems
(\ref{3-6}) and (\ref{3-8}) using the conditional symmetries
 obtained in Section \ref{sec-4}, examine their properties and suggest a possible interpretation.
 We note that the case $S=1$ is not examined because the linear terms $u$ and $v$
 are removable in this case (see (\ref{2-2})), so that the real-world applicability
 of the system obtained are questionable.

 Let us consider \textbf{Case I} of Theorem~\ref{th-3}. To
 find exact solutions of the DHT system (\ref{3-6}) using the
 operator $Q^u_1$, we
 apply the standard procedure. Firstly,  the invariant surface conditions
  \[ Q^u_1(u)=0, \ Q^u_1(v)=0, \]
  i.e.
   \be\label{4-1}
  u_x=f(t)u, \quad v_x=f(t)v-\frac{f'(t)}{S}\,u
   \ee
 should be solved.  (\ref{4-1}) is  the system of
the linear first-order PDEs involving the function $f(t)$ in an
implicit form (see  (\ref{3-7})). Solving this  system, one obtains
the ansatz \be\label{4-2}\ba{l}
u(t,x)=\varphi(t)e^{xf(t)}, \medskip\\
v(t,x)=\psi(t)e^{xf(t)}-\frac{xf'}{S}\,\varphi(t)e^{xf(t)}, \ea\ee
where $\varphi$ and $\psi$ are new unknown  functions. Substituting
ansatz (\ref{4-2}) into (\ref{3-6}) and taking into account
(\ref{3-7*}),
 one derives  the reduced system of
ODEs \be\label{4-3}\ba{l}
\varphi'=\lf(1+f^2\rg)\varphi-S\psi, \medskip\\
\psi'=\lf(S+df^2\rg)\psi-\frac{S}{\varphi}\psi^2-\frac{2dff'}{S}\varphi.
\ea\ee System  (\ref{4-3})  can be rewritten in the form
\be\label{4-4}\ba{l}\psi=\frac{\varphi'-\lf(1+f^2\rg)\varphi}{S}, \medskip\\
\varphi''-\frac{{\varphi'}^2}{\varphi}+K_1(t)\varphi'+K_0(t)\varphi=0,
\ea\ee where \be\label{4-6}K_1(t)=1-S+(1-d)f^2, \
K_0(t)=S-1+(d+S-2)f^2+(d-1)f^4-2(d+1)ff'.\ee
 By applying  substitution (\ref{2-23}), the second
 equation of system (\ref{4-4}) reduces to the linear ODE
\be\label{4-5} \chi'+K_1\chi+K_0=0.\ee Thus, solving the  equations
(\ref{4-5}) and (\ref{2-23})
and taking into account formulae (\ref{4-4}) and (\ref{4-6}), one
obtains the functions $\varphi$ and $\psi$. Finally, exact solutions
of the DHT system (\ref{3-6}) in  the form (\ref{4-2}) can be easily
written down.  From the applicability point of view, the solutions
obtained are not convenient because they involve  the function
$f(t)$ in an implicit form.

 Now we present an example constructing the exact solution on the DHT
system (\ref{3-6}) in an explicit form. Firstly, one
 needs  to identify the function $f$ in  the explicit form. Obviously, it occurs if one sets $C_1=0$  in (\ref{3-7}).  Note that an arbitrary  constant  $C_2$ can be taken
  $C_2=0$ as well  without loosing a generality. In this special case
 \be\label{4-7}f=\pm \sqrt{\frac{S-1}{1+\frac{1-d}{3}\,e^{2\lf(S-1\rg)t}}}\,e^{\lf(S-1\rg)t}.\ee
Thus, the  coefficients $K_0$ and $K_1$  in (\ref{4-5}) have the
forms \be\label{4-8}\ba{l}
K_1=\frac{(1-S)\,\lf(3+2(d-1)e^{2\lf(S-1\rg)t}\rg)}{3+(1-d)e^{2\lf(S-1\rg)t}},\medskip\\
K_0=\frac{1-S}{\lf(3+(1-d)e^{2\lf(S-1\rg)t}\rg)^2}\,\lf(b_2e^{4\lf(S-1\rg)t}+b_1e^{2\lf(S-1\rg)t}-9\rg),
\ea\ee where
\[ b_1=3(3S+6dS-7d-2), \quad  b_2= 2(d-1)(2+d-3S). \]

After straightforward calculations, we obtain the following
functions $\varphi$ and $\psi$\,:
 \be\label{4-9}\ba{l}
\varphi(t)=\exp\lf(G(t)+t\rg)\lf(3+(1-d)e^{2\lf(S-1\rg)t}\rg)^{\frac{3(2d-1)}{2(d-1)}},\medskip\\
\psi(t)=\frac{3(1-S)\,G(t)}{S\lf(3+(1-d)\,e^{2\lf(S-1\rg)t}\rg)}\,\varphi(t),\ea\ee
where
\[G(t)=
\left\{ \begin{array}{l}
\frac{e^{\lf(S-1\rg)t}}{\sqrt{3+(1-d)e^{2\lf(S-1\rg)t}}}\lf[C+\frac{6d}{\sqrt{1-d}}
arcsinh\lf(\frac{\sqrt{1-d}}{\sqrt{3}}e^{\lf(S-1\rg)t}\rg)\rg], \quad d<1,\\
\frac{e^{\lf(S-1\rg)t}}{\sqrt{3+(1-d)e^{2\lf(S-1\rg)t}}}\lf[C+\frac{6d}{\sqrt{d-1}}
arcsin\lf(\frac{\sqrt{d-1}}{\sqrt{3}}e^{\lf(S-1\rg)t}\rg)\rg], \quad
d>1.
\end{array} \right.
\]
Formulae (\ref{4-7}) and (\ref{4-9}) lead to the exact solution
\[\ba{l} u(t,x)=\Big(3+
(1-d)e^{2\lf(S-1\rg)t}\Big)^{\frac{3(2d-1)}{2(d-1)}}\exp\lf(G(t)+t\pm\sqrt{\frac{3(S-1)}{3+(1-d)e^{2\lf(S-1\rg)t}}}\,e^{\lf(S-1\rg)t}\,x\rg),\medskip\\
v(t,x)=\lf[\frac{3(1-S)}{S}\Big(3+(1-d)\,e^{2\lf(S-1\rg)t}\Big)^{-1}\,G(t)\mp
\sqrt[3]{\frac{3(S-1)}{3+(1-d)e^{2\lf(S-1\rg)t}}}e^{(S-1)t}\,x\rg]\,u(t,x)
\ea\] of the DHT system (\ref{3-6}).

\textbf{Case II} of Theorem~\ref{th-3}. Solving the first-order PDEs
 \[ Q^u_2(u)=0, \ Q^u_2 (v)=0, \]  one
obtains the ansatz \be\label{4-12}\ba{l}
u(t,x)=\varphi(t)\exp\lf(\frac{h}{g}\,x-\frac{g'}{4g}\,x^2\rg), \medskip\\
v(t,x)=\lf(\psi(t)-\frac{\varphi(t)\,x}{4Sg}\lf(4h'-g''x\rg)\rg)\exp\lf(\frac{h}{g}\,x-\frac{g'}{4g}\,x^2\rg),
\ea\ee where $\varphi$ and $\psi$ are functions to be found, while
the functions $g$ and $h$ satisfy system (\ref{3-9}). Substituting
ansatz (\ref{4-12}) into the DHT system (\ref{3-8}), we arrive at
the reduced system of ODEs \be\label{4-13}\ba{l}
\varphi'=\lf(1+\frac{h^2}{g^2}-\frac{g'}{2g}\rg)\varphi-S\psi, \medskip\\
\psi'=\lf(S+\frac{h^2}{g^2}-\frac{g'}{2g}\rg)\psi-\frac{S}{\varphi}\psi^2+\frac{gg''-4hh'}{2Sg^2}\varphi.
\ea\ee

Similarly to the ODE system  (\ref{4-4}),  system  (\ref{4-13})  can
be easily solved. Indeed, using  the first equation of system
(\ref{4-13}), we immediately obtain
\[\psi=\frac{1}{S}\lf[-\varphi'+\lf(1+\frac{h^2}{g^2}-\frac{g'}{2g}\rg)\varphi\rg].\]
Substituting the function $\psi$ and its first-order derivative into
the second equation of (\ref{4-13}) and applying  the substitution
(\ref{2-23}), we obtain the linear ODE
\be\label{4-19}\chi'+(1-S)\chi+K(t)=0,\ee where
\[K(t)=S-1+\frac{1}{2g^2}\Big(2g^2g''-g{g'}^2+(4h^2+g^2-Sg^2)g'+2h(Sh-h-4h')g\Big).\]
Thus, having the  correctly-specified  functions  $g$ and $h$, ODE
(\ref{4-19}) can be easily solved and the functions $\varphi$ and
$\psi$ constructed. To obtain exact solutions of the DHT
system~(\ref{3-8}) in an explicit form,   we use the functions $g$
and $h$ from  (\ref{3-10}). Thus,
\[K(t)=-\frac{S^2-10S+9}{8}-\frac{4C_3(C_2+C_3t)}{C_1^2}.\]
Integrating (\ref{4-19}) and (\ref{2-23}) and renaming $C_2
\rightarrow C_1\lf(C_2-2C_3\rg), \ C_3\rightarrow
\left(S-1\rg)C_1C_3$, we arrive at the solution
\be\label{4-14}\ba{l}
\varphi(t)=\exp\Big[\frac{9-S}{8}\,t-4C_3(C_2-C_3)\,t+2C_3^2\left(1-S\rg)\,t^2+Ce^{(S-1)t}\Big], \medskip\\
\psi(t)=\frac{1}{S}\Big[\frac{1-S}{8}+\big(C_2+C_3(S-1)t\big)^2+(1-S)Ce^{(S-1)t}\Big]\varphi(t)
\ea\ee of the ODE system (\ref{4-13}).

Finally, substituting (\ref{4-14}) into ansatz (\ref{4-12}) and
taking into account (\ref{3-10}), we obtain the three-parameter
family of  exact solutions of system (\ref{3-8})
 \begin{small}\be\label{4-20}\ba{l}
u(t,x)=\exp\Big[\lf(\frac{9-S}{8}-4C_3^2\rg)\,t+(2C_3-C_2)(4C_3t-x)
-\frac{1}{8}(S-1)(4C_3t-x)^2+Ce^{(S-1)t}\Big],\\
\medskip\\
v(t,x)=\frac{1}{16S}\Big[\big(4C_2+(S-1)(4C_3t-x)\big)^2-2(S-1)\big(1+8Ce^{(S-1)t}\big)\Big]u(t,x),
\ea\ee \end{small}where $C, \, C_2$ and $ C_3$ are arbitrary
parameters.

\begin{Remark} \label{rem4}Since the DHT system (\ref{3-8}) admits the Galilei operator $G$,
the exact solution (\ref{4-20}) can be presented in the simpler
form. Indeed,using formula  (\ref{2-25}), solution
 (\ref{4-20}) can be simplified  to the form  \[\ba{l}
u(t,x)=\exp\Big[\frac{9-S}{8}\,t+C_2x
-\frac{S-1}{8}x^2+Ce^{(S-1)t}\Big],\\
\medskip\\
v(t,x)=\frac{1}{16S}\Big[\big(4C_2-(S-1)x\big)^2-2(S-1)\big(1+8Ce^{(S-1)t}\big)\Big]u(t,x)
\ea\] (here the upper index $*$ is omitted).  The relevant  Galilei
transformation is
\[x^*=x-4C_3t, \ u^*=\exp\lf(2C_3x-4C_3^2t\rg)u, \ v^*=\exp\lf(2C_3x-4C_3^2t\rg)v.\]
\end{Remark}

To present some biological interpretation, one may consider a
specific solution by setting $C_2=C_3=0$ in (\ref{4-20}). In this
case,  the exact solution
is as follows  \be\label{4-16}\ba{l}
u(t,x)=\exp\lf(\frac{9-S}{8}\,t+Ce^{(S-1)t}-\frac{S-1}{8}\,x^2\rg), \medskip\\
v(t,x)=\frac{S-1}{16S}\lf(-2-16Ce^{(S-1)t}+(S-1)x^2\rg)u(t,x).
\ea\ee

\begin{figure}[h!]
\begin{center}
\includegraphics[width=6cm]{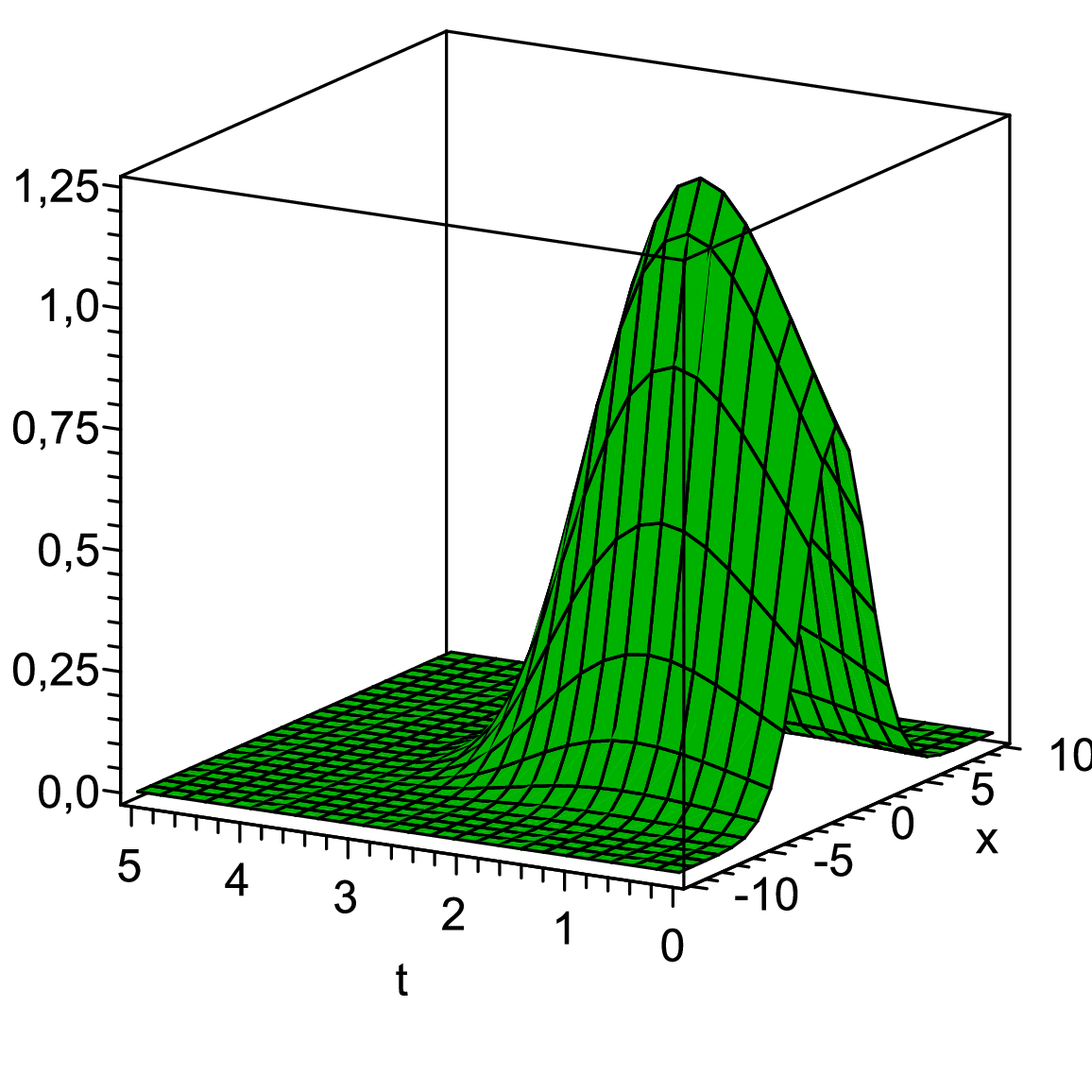}
\includegraphics[width=6cm]{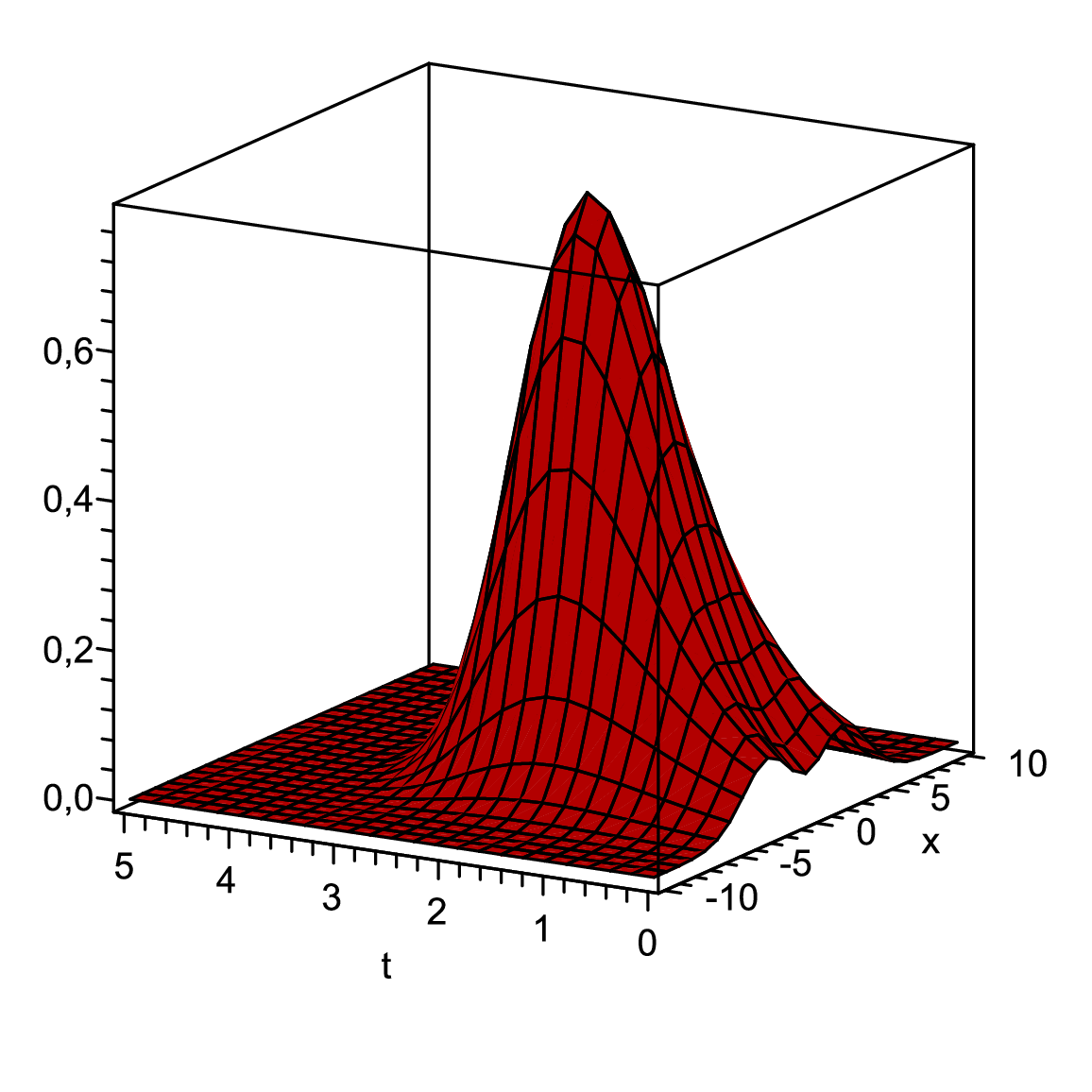}
\end{center}
\caption{Surfaces representing the $u$ (green) and $v$ (red)
components  of the exact  solution (\ref{4-16}) of the DHT system
(\ref{3-8}). The parameters $S=2, \ C=-0.25$. } \label{f3}
\end{figure}

\begin{figure}[h!]
\begin{center}
\includegraphics[width=6cm]{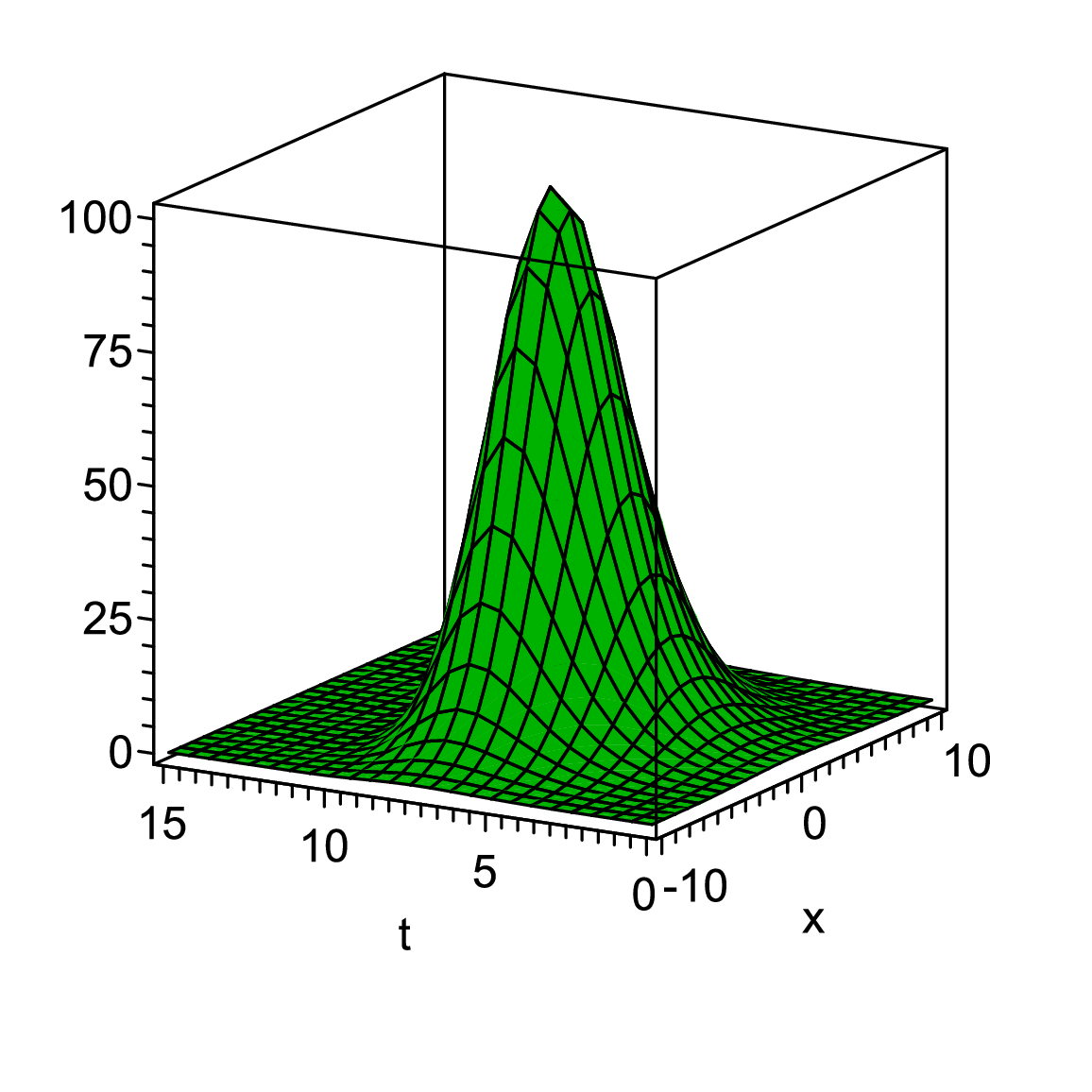}
\includegraphics[width=6cm]{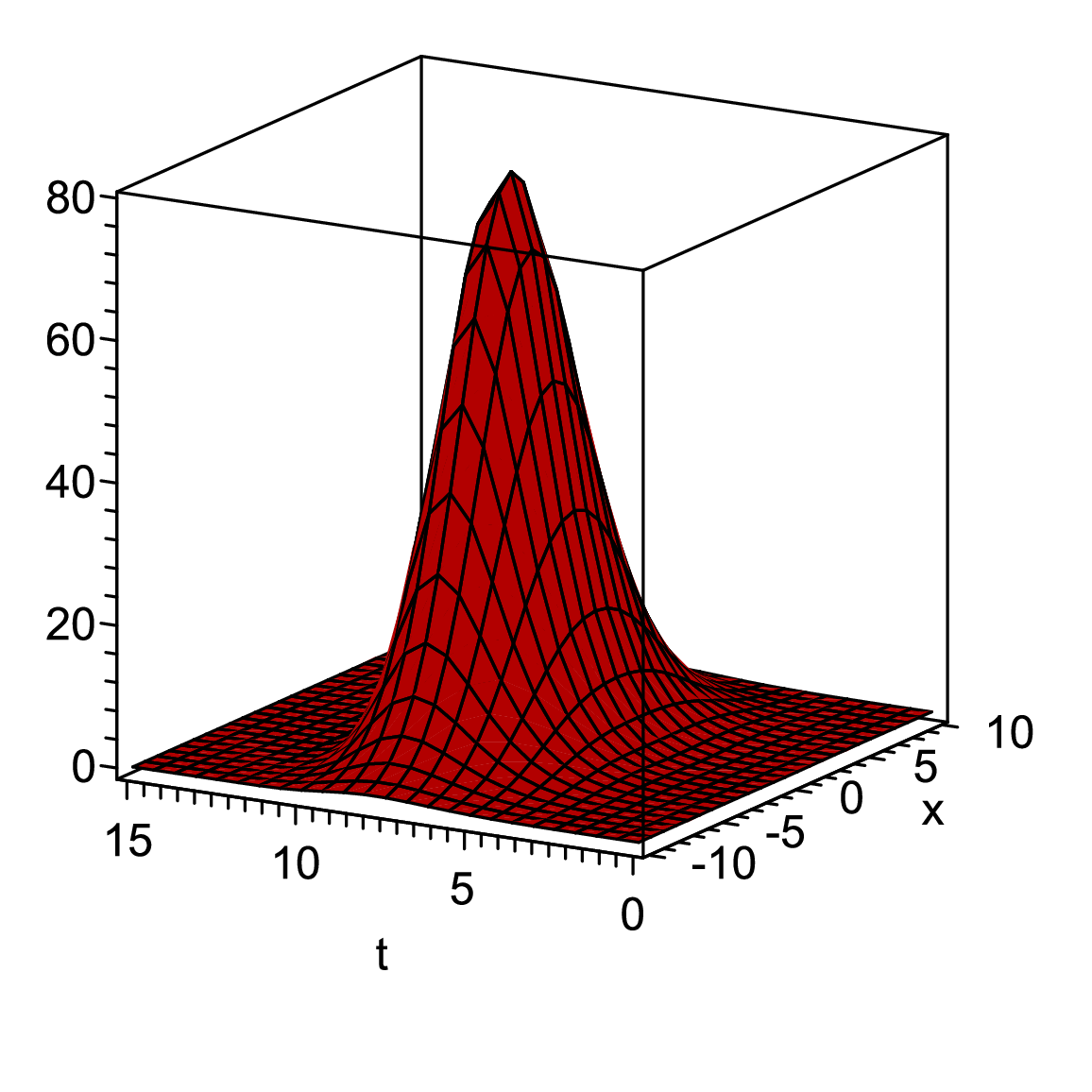}
\end{center}
\caption{Surfaces representing the $u$ (green) and $v$ (red)
components  of the exact  solution (\ref{4-16}) of the DHT system
(\ref{3-8}). The parameters $S=1.4, \ C=-0.125$. } \label{f4}
\end{figure}

The components of solution (\ref{4-16}) are bounded and positive in
the domain \[\Omega=\left\{ (t,x) \in (0,+ \infty)\times (- \infty,
+ \infty)\rg\}\] provided the coefficient restrictions
$C\leq-\frac{1}{8}, \ S>1$ hold. Moreover, solution (\ref{4-16})
satisfies the zero Neumann conditions at infinity \[
 u_x(t, -\infty)=u_x(t, +\infty)= v_x(t, -\infty)=v_x(t, +\infty)=0
 \]
in the domain $\Omega$. One also easily checks  that  the asymptotic
behaviour
\[ (u,\,v)  \rightarrow  (0,\,0) \quad \mbox{as} \quad
t  \rightarrow +\infty \] takes place, which predicts a total
extinction of the both species $u$ and $v$. An example of solution
(\ref{4-16}) is presented in Fig.~\ref{f3} (Fig.~\ref{f4} highlights
the same scenario for preys and predators but with the initial
profiles approaching zero densities). One can note that the
zero-flux conditions are satisfied with a sufficient exactness at
any bounded space interval $[a, \ b]$ with $ |a|>10,  \ |b|>10$. It
can be regarded as a
 habit of the both species to concentrate themselves in a compact area. For example,
  it may happen if the compact area is a forest with large amount of food for preys.
  To the best of our knowledge, such type solutions are unknown for the diffusive
  Lotka--Volterra system modelling the prey-predator interaction.

\begin{figure}[h!]
\begin{center}
\includegraphics[width=12cm]{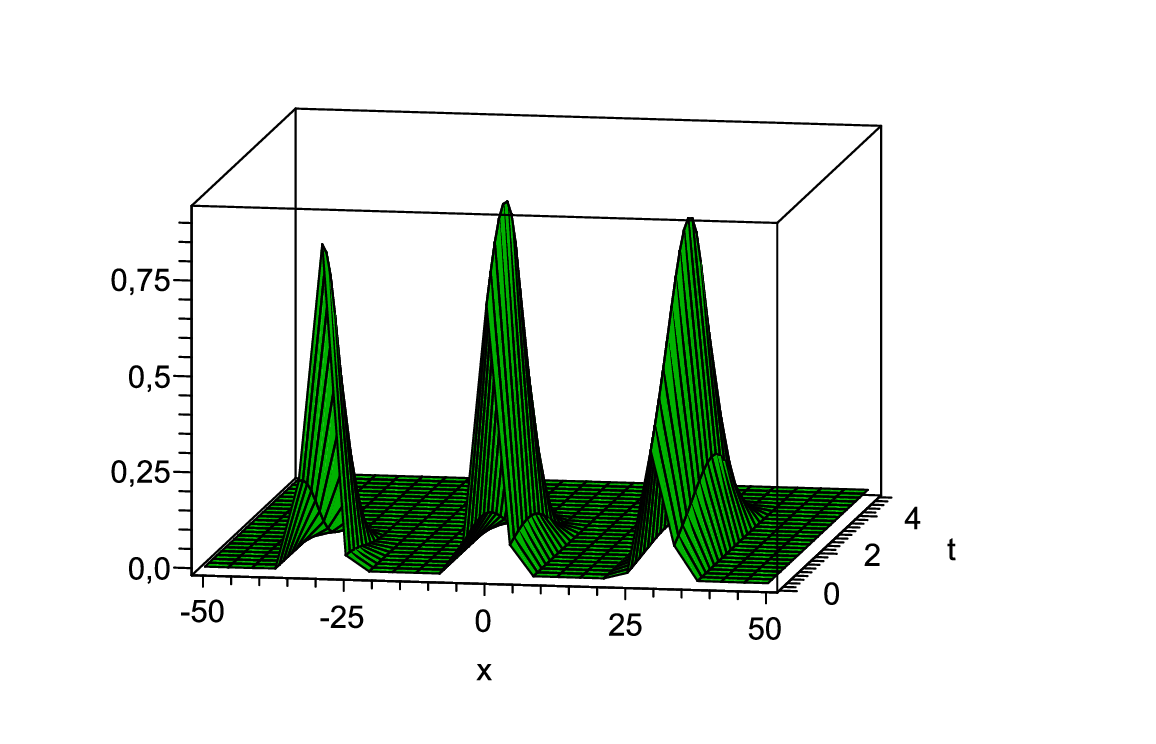}
\end{center}
\caption{Surface representing the component $U$ of the approximate
solution (\ref{4-20}) of the DHT systems (\ref{3-8}). The parameters
$S=2, \ C=-0.35$, $x_0=-30, \ x_1=0, \ x_2=30, \ t_0=-1, \ t_1=0, \
t_2=1.$ } \label{f5}
\end{figure}
\begin{figure}[h!]
\begin{center}
\includegraphics[width=12cm]{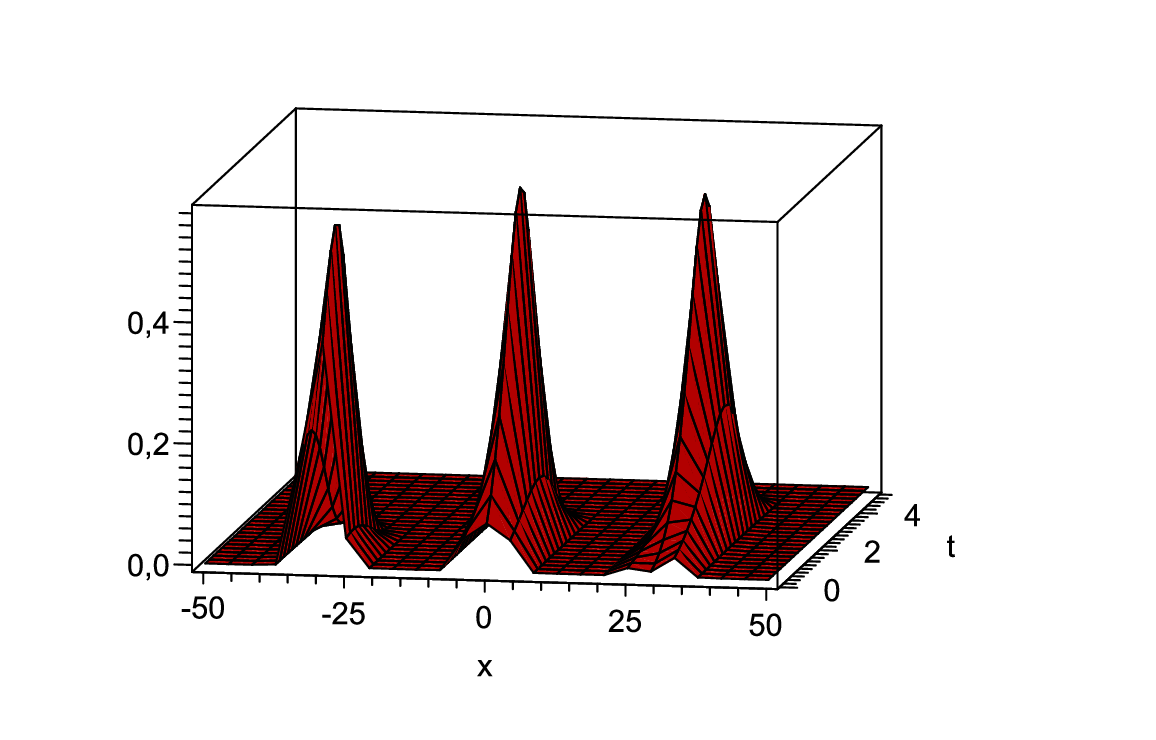}
\end{center}
\caption{Surface representing the component $V$ of the approximate
solution (\ref{4-20}) of the DHT systems (\ref{3-8}). The parameters
$S=2, \ C=-0.35$,  $x_0=-30, \ x_1=0, \ x_2=30, \ t_0=-1, \ t_1=0, \
t_2=1.$} \label{f6}
\end{figure}

Now we present an observation, which is highly unusual for nonlinear
PDEs, for which the
 well-known principle of linear superposition of solutions is not valid.
Because the DHT  model admits the time and space translations,
the exact solution (\ref{4-16}) can be rewritten in the form
 \be\label{4-16*}\ba{l}
u(t,x)=\exp\Big[\frac{9-S}{8}\,(t+t_0)+C\exp\big((S-1)(t+t_0)\big)-\frac{S-1}{8}\,(x+x_0)^2\Big], \medskip\\
v(t,x)=\frac{S-1}{16S}\Big[-2-16C\exp\big((S-1)(t+t_0)\big)+(S-1)(x+x_0)^2\Big]u(t,x),
\ea\ee where $t_0$ and $x_0$ are arbitrary parameters. Similarly to
(\ref{4-16}), each exact solution (\ref{4-16*}) with a fixed pair
$(t_0, x_0)$  is bounded and positive in the domain $\Omega$
provided  the restrictions $C\leq C_0 =-\frac{1}{8}e^{(1-S)t_0}, \
S>1$ hold. On the other hand, the solution decays very sharply if
$|x+x_0| \to \infty$, because both components contain multiplier
$\exp\lf(-\frac{S-1}{8}\,(x+x_0)^2\rg)$.
 For example, one notes  from Fig.~\ref{f3} that both components $u$ and $v$ of solution (\ref{4-16}) practically vanish if $|x|>10$. So, taking solution (\ref{4-16*}), say with
  $t_0=0$ and $x_0=30$, one observes that this solution vanishes  with a high exactness  beyond the interval $x \in [-40,-20]$. Now we conclude that the expressions
  \[ u(t,x)+u(t+t_0,x+x_0),  \quad v(t,x)+v(t+t_0,x+x_0), \]
  where the functions  $u$ and $v$ are the exact solutions of  the form (\ref{4-16}) and (\ref{4-16*}),
  produce an approximate solution  of the DHT
systems (\ref{3-8}) provided $|x_0|$ is sufficiently large.
Moreover, a further generalization of the form \be\label{4-21}\ba{l}
U(t,x)= u(t+t_0,x+x_0)+u(t+t_1,x+x_1)+\dots+ u(t+t_m,x+x_m), \\
V(t,x)= v(t+t_0,x+x_0)+v(t+t_1,x+x_1)+\dots+ v(t+t_m,x+x_m) \ea\ee
presents another  approximate solution  of the DHT systems
(\ref{3-8}) provided the differences $|x_i-x_j|, \ (0\leq i<j\leq
m)$ are sufficiently large. An example of the approximate  solution
(\ref{4-21}) is presented in Fig.\,\ref{f5}--Fig.\,\ref{f6}.

\section{Conclusions} \label{sec-6}

In this work, the DHT prey-predator model (under the assumption of
the Malthusian growth low for preys)   is investigated by means of
Lie and $Q$-conditional (nonclassical) symmetry methods.
 Applying the classical  Lie method, all Lie symmetries of the model in question were found (see Theorem~\ref{th-1})
and it was shown that these symmetries coincide with those, which
follow from the earlier works  although  \cite{ch-king1,ch-king2}.
The results presented in Theorems \ref{th-2} and \ref{th-3} are new
and highly nontrivial. In fact, $Q$-conditional symmetries of the
DHT system (\ref{1-1}) are found for the first time. It is also
proved that this system admits only those $Q$-conditional symmetries
with  $\xi^0\not=0$ (see (\ref{3-2})), which are equivalent to the
Lie symmetry operators listed in Theorem~\ref{th-1}.  However,
applying the algorithm based on the recently introduced
Definition~\ref{d2}, we found new $Q$-conditional symmetries in the
`no-go case' ($\xi^0=0$ in (\ref{3-2})).

Both Lie and $Q$-conditional symmetries are applied for construction
of exact solutions via reduction of  the DHT system (\ref{1-1}) to
ODE systems and solving the latter.
 The most interesting (from the
applicability point of view) Cases 1 and 3 of Table~\ref{tab1}  has
been examined.  As a result, several families of exact solutions are
derived. Interestingly, the exact solutions derived via
$Q$-conditional symmetries (see formulae
(\ref{4-16})--(\ref{4-16*})) are non-Lie solutions, i.e. they cannot
be found using the Lie ans\"atze listed in Table \ref{tab2}. We
remind the reader that a given $Q$-conditional symmetry, generally
speaking, does not lead to  non-Lie solutions (see the relevant
discussion and examples in Chapter 4 \cite{ch-se-pl-book}).

Because some solutions (see (\ref{2-34*}) and  (\ref{4-16})) possess
remarkable properties, in particular
 boundedness and nonnegativity, a possible biological interpretation is suggested and the relevant plots are presented.

   Finally, we presented  a formula for  generation of approximate solutions of the DHT system (\ref{1-1}).
   The formula allows us to construct such approximate solutions, which
   consist of several `peaks' (see Fig.\,\ref{f5} and \ref{f6}). Interestingly that the approximate solutions  (\ref{4-21})
    have a very  similar structure to the numerical solutions derived in \cite{ni-98} (see Fig.\,3 therein) for the
activator-inhibitor model \be\label{act-inhib}\ba{l}
u_t = d_1u_{xx} -a_1u + b_1v^2, \medskip\\
v_t = d_2v_{xx} -a_2v + b_2\frac{v^2}{u}  \ea\ee (here all
parameters are positive constants). System  (\ref{act-inhib}) was
developed in the seminal work \cite{gierer-mein-72} and is motivated
by biological experiments on hydra. One easily notes that the above
system coincides with the DHT sytem (\ref{2-11}) if one replace the
quadratic term $b_1v^2$ by  linear  and takes the opposite signs in
reactive terms.

\emph{\textbf{Acknowledgments.}} R.Ch. acknowledges that this
research was funded by the British Academy's Researchers at Risk
Fellowships Programme. The authors are grateful to Reviewer~1 for
fruitful comments that can be used in future studies of the DHT
system.


\begin{thebibliography}{99}


\bibitem{arancibia-21} Arancibia-Ibarra, C., Bode, M., Flores, J.,
Pettet, G.,  Van Heijster, P.:  Turing patterns in a diffusive
Holling--Tanner predator-prey model with an alternative food source
for the predator. Comm. Nonlinear Sci. Numer. Simulat. \textbf{99},
105802 (2021)

\bibitem {aris-75I} Aris, R.:
 The Mathematical Theory of Diffusion and
 Reaction in Permeable Catalysts: the Theory of
  the Steady State. Clarendon Press, Oxford (1975)

\bibitem{arrigo-15} Arrigo, D.J.: Symmetry Analysis of Differential Equations: an
Introduction. John Wiley \& Sons, Hoboken (2015)

\bibitem{alaoui-18}  Aziz-Alaoui, M.A.,  Daher Okiye, M., Moussaoui, A.: Permanence
and extinction of a diffusive predator-prey model with Robin
boundary conditions. Acta Biotheoretica \textbf{66}, 367--378 (2018)

\bibitem{bl-anco-10}  Bluman, G.W., Cheviakov, A.F., Anco, S.C.: Applications of Symmetry
Methods to Partial Differential Equations. Springer, New York (2010)

\bibitem {britton}   Britton, N.F.:  Essential Mathematical
Biology. Springer, Berlin (2003)

\bibitem{bro-ch-goa-22} Broadbridge, P.,  Cherniha, R.M.,  Goard,
J.M.: Exact nonclassical symmetry solutions of Lotka--Volterra-type
population systems. Euro. J. Appl. Math. 1--19, (2022)

\bibitem{chen-12} Chen, S., Shi, J.: Global stability in a diffusive
Holling--Tanner predator-prey model. Appl. Math. Lett. \textbf{25},
614--618 (2012)

\bibitem{ch-88}  Cherniha, R.M.: On exact solutions of a nonlinear
 diffusion-type system.
 In: Symmetry analysis and exact solutions of
  equations of mathematical physics, Kyiv, Inst. Math. Ukrainian
Acad. Sci., 49--53 (1988)

\bibitem{ch-2010}  Cherniha, R.:  Conditional symmetries
 for systems of PDEs:  new definition and their application for
 reaction-diffusion systems.  J. Phys. A: Math.
 Theor.  \textbf{43}, 405207 (2010)


\bibitem{ch-dav-book}  Cherniha, R.,  Davydovych, V.:
Nonlinear Reaction-Diffusion Systems --- Conditional Symmetry, Exact
Solutions and their Applications in Biology.  Lecture Notes in
Mathematics, vol. 2196. Springer, Cham (2017)


\bibitem{ch-dav-math-21}  Cherniha, R., Davydovych,
V.: New conditional symmetries and exact solutions of the diffusive
two-component Lotka--Volterra system.  Mathematics \textbf{9}, 1984
(2021)

\bibitem{ch-dav-AAM-22} Cherniha, R.,  Davydovych, V.: A hunter--gatherer-farmer population model: new
conditional symmetries and exact solutions with biological
interpretation. Acta Appl. Math. \textbf{182}, 4 (2022)

  \bibitem {ch-king1}  Cherniha, R.,  King, J.R.:
 Lie   symmetries of nonlinear  multidimensional
reaction-diffusion systems: I. J. Phys A: Math. Gen. \textbf{33},
267--282 (2000)

\bibitem {ch-king2} Cherniha, R.,  King, J.R.:
 Lie   symmetries of nonlinear  multidimensional
reaction-diffusion systems: II. J. Phys A: Math. Gen. \textbf{36},
405--425 (2003)

\bibitem {ch-se-pl-book}  Cherniha, R., Serov, M.,  Pliukhin, O.:
Nonlinear Reaction-Diffusion-Convection Equations: Lie and
Conditional Symmetry, Exact Solutions and their Applications.
Chapman and Hall/CRC, New York  (2018)

\bibitem {fife-79} Fife, P.:  Mathematical Aspects of Reacting and Diffusing Systems.
Springer, New York (1975)

\bibitem{fu-ch-95} Fushchych, W.I., Cherniha, R.M.: Galilei-invariant systems of nonlinear systems of
evolution equations. J. Phys. A: Math. Gen. \textbf{28}, 5569--5579
(1995)

\bibitem {gierer-mein-72} Gierer, A., Meinhardt, H.: A theory of biological pattern formation.  Kybernetik
\textbf{12}, 30--39 (1972)

\bibitem{hanski91} Hanski, I., Hansson, L., Henttonen, H.: Specialist
predators, generalist predators, and the microtine rodent cycle. J.
Anim. Ecol. \textbf{60}, 353--367 (1991)

\bibitem{hanski93} Hanski, I., Turchin, P., Korpim\"{a}ki, E., Henttonen,
H.: Population oscillations of boreal rodents: regulation by
mustelid predators leads to chaos. Nature \textbf{364}, 232--235
(1993)

\bibitem {ku-na-ei-16} Kuang, Y., Nagy, J.D., Eikenberry, S.E.: Introduction to Mathematical Oncology.
CRC Press, Boca Raton (2016)

\bibitem{ji-16} Lina, J.i.:
 The method of linear determining equations to evolution
  system and application for reaction-diffusion system with power diffusivities.
Symmetry \textbf{8}, 157 (2016)

\bibitem{lot}  Lotka, A.J.:
Undamped oscillations derived from the law of mass action. J. Am.
Chem. Soc. \textbf{42}, 1595--1599 (1920)

\bibitem {mur2}  Murray, J.D.: Mathematical Biology. Springer, Berlin
(1989)

\bibitem {mur2002} Murray, J.D.:
Mathematical Biology I. Springer, Berlin (2002)

\bibitem {mur2003} Murray, J.D.:
Mathematical Biology II. VSpringer, Berlin (2003)

\bibitem {ni-98} Ni, W.M.: Diffusion, cross-diffusion,
and their spike-layer steady states. Not. AMS \textbf{45}, 9--18
(1998)

\bibitem {okubo} Okubo, A., Levin, S.A.: Diffusion and Ecological Problems.
Modern Perspectives, 2nd edn. Springer, Berlin (2001)

\bibitem {olv-93}  Olver, P.:
Applications of Lie Groups to Differential Equations. 2nd edn.
Springer, Berlin (1993)

\bibitem{Pat-Wint-77} Patera, J.,  Winternitz, P.: Subalgebras of real three- and
four-dimensional Lie algebras. J. Math. Phys. \textbf{18},
1449--1455 (1977)

\bibitem{Qi-16} Qi, Y., Zhu, Y.:  The study of global stability of
 a diffusive Holling--Tanner predator-prey model. Appl. Math. Lett. \textbf{57},
 132--138 (2016)


\bibitem{tanner} Tanner, J.T.  The stability and the intrinsic growth rates
of prey and predator populations. Ecology \textbf{56}, 855--867
(1975)

\bibitem{torrisi-21} Torrisi, M., Tracina, R.: Lie
symmetries and solutions of reaction-diffusion systems arising in
biomathematics. Symmetry \textbf{13}, 1530 (2021)

\bibitem{torrisi-23} Torrisi, M., Tracina, R.:
Symmetries and solutions for a class of ddvective reaction-diffusion
systems with a special reaction term. Mathematics \textbf{11}, 160
(2023)

\bibitem{vol}  Volterra, V.:  Variazioni e fluttuazioni del numero
d`individui in specie animali conviventi. Mem. Acad. Lincei.
\textbf{2}, 31--113 (1926)


\bibitem{wollkind} Wollkind, D.J., Collings, J.B., Logan, J.A.:
Metastability in a temperature-dependent model system for
predator-prey mite outbreak interactions on fruit trees. Bull. Math.
Biol. \textbf{50}, 379--409 (1988)


 \bibitem {zhdanov}  Zhdanov, R.Z., Lahno, V.I.:  Conditional symmetry of a porous
medium equation.   Phys. D \textbf{122}, 178--86 (1998)

\end{thebibliography}
\end{document}